\documentclass[twocolumn]{aastex631}
\UseRawInputEncoding

\newcommand\hxmt{{\it Insight}-HXMT}
\newcommand\target{Swift J1727.8--1613}

\submitjournal{ApJL}

\begin{document}

\title{A timing view of the additional high-energy spectral component discovered in the black hole candidate Swift J1727.8--1613}

\correspondingauthor{Zi-Xu Yang}
\email{yangzx@sdut.edu.cn}

\correspondingauthor{Liang Zhang}
\email{zhangliang@ihep.ac.cn}

\author[0000-0003-1718-8487]{Zi-Xu Yang}
\affiliation{School of Physics and Optoelectronic Engineering, Shandong University of Technology, Zibo 255000, China}

\author[0000-0003-4498-9925]{Liang Zhang}
\affiliation{Key Laboratory of Particle Astrophysics, Institute of High Energy Physics, Chinese Academy of Sciences, Beijing 100049, China}

\author[0000-0001-5586-1017]{Shuang-Nan Zhang}
\affiliation{Key Laboratory of Particle Astrophysics, Institute of High Energy Physics, Chinese Academy of Sciences, Beijing 100049, China}

\author[0000-0002-2705-4338]{Lian Tao}
\affiliation{Key Laboratory of Particle Astrophysics, Institute of High Energy Physics, Chinese Academy of Sciences, Beijing 100049, China}

\author{Shu Zhang}
\affiliation{Key Laboratory of Particle Astrophysics, Institute of High Energy Physics, Chinese Academy of Sciences, Beijing 100049, China}

\author[0000-0003-3260-8718]{Ruican Ma}
\affiliation{Key Laboratory of Particle Astrophysics, Institute of High Energy Physics, Chinese Academy of Sciences, Beijing 100049, China}

\author[0000-0001-5238-3988]{Qing-Cui Bu}
\affil{Institut f\"ur Astronomie und Astrophysik, Kepler Center for Astro and Particle Physics, Eberhard Karls Universit\"at, Sand 1, 72076 T\"ubingen, Germany」}

\author[0000-0002-3515-9500]{Yue Huang}
\affiliation{Key Laboratory of Particle Astrophysics, Institute of High Energy Physics, Chinese Academy of Sciences, Beijing 100049, China}

\author[0000-0002-8032-7024]{He-Xin Liu}
\affiliation{Key Laboratory of Particle Astrophysics, Institute of High Energy Physics, Chinese Academy of Sciences, Beijing 100049, China}

\author{Wei Yu}
\affiliation{Key Laboratory of Particle Astrophysics, Institute of High Energy Physics, Chinese Academy of Sciences, Beijing 100049, China}

\author[0000-0002-3718-9252]{Guangcheng Xiao}
\affiliation{ Department of Physics, Jinggangshan University, Jiangxi Province, Ji'an 343009, People's Republic of China}

\author{Peng-Ju Wang}
\affiliation{Key Laboratory of Particle Astrophysics, Institute of High Energy Physics, Chinese Academy of Sciences, Beijing 100049, China}

\author{Hua Feng}
\affiliation{Key Laboratory of Particle Astrophysics, Institute of High Energy Physics, Chinese Academy of Sciences, Beijing 100049, China}

\author[0000-0003-0274-3396]{Li-Ming Song}
\affiliation{Key Laboratory of Particle Astrophysics, Institute of High Energy Physics, Chinese Academy of Sciences, Beijing 100049, China}

\author[0000-0002-2032-2440
]{Xiang Ma}
\affiliation{Key Laboratory of Particle Astrophysics, Institute of High Energy Physics, Chinese Academy of Sciences, Beijing 100049, China}

\author[0000-0002-3776-4536]{Mingyu Ge}
\affiliation{Key Laboratory of Particle Astrophysics, Institute of High Energy Physics, Chinese Academy of Sciences, Beijing 100049, China}

\author[0000-0001-9893-8248]{Qing-Chang Zhao}
\affiliation{Key Laboratory of Particle Astrophysics, Institute of High Energy Physics, Chinese Academy of Sciences, Beijing 100049, China}

\author[0000-0002-9796-2585]{Jin-Lu Qu}
\affiliation{Key Laboratory of Particle Astrophysics, Institute of High Energy Physics, Chinese Academy of Sciences, Beijing 100049, China}




\begin{abstract}
We present an energy-dependent analysis for the type-C quasi-periodic oscillations (QPOs) observed in the black hole X-ray binary \target\ using \hxmt\ observations. We find that the QPO fractional rms at energies above 40 keV is significantly higher than that below 20 keV. This is the first report of a high energy (HE)-rms excess in the rms spectrum of a black hole X-ray binary. 
In the high energy band, an extra hard component is observed in additional to the standard thermal Comptonization component at similar energy band. The value of the QPO HE-rms excess is not only correlated with the disk parameters and the photon index of the standard Comptonization component, but also exhibits a moderate positive correlation with the flux of the additional hard spectral component. No features in the QPO phase-lag spectra are seen corresponding to the additional hard component. We propose that the additional hard component in the spectrum may originate from jet emission and the associated QPO HE-rms excess can be explained by the precession of the jet base.

\end{abstract}



\section{Introduction} 

Black hole low-mass X-ray binaries (BH-LMXBs) are transient sources that exhibit distinct spectral-timing states during an outburst. Four main states can be identified based on their spectral-timing characteristics  \citep[see][for recent reviews]{2011BASI...39..409B,2016ASSL..440...61B}: the Hard state (HS), Hard-Intermediate state (HIMS), Soft-Intermediate state (SIMS) and Soft state (SS). The spectra of the HS are dominated by a thermal Comptonization component from a hot corona/jet base, while the SS spectra are typically dominated by a soft component from an accretion disc.
Type-C low-frequency QPOs with frequencies ranging from 0.1 Hz to 10 Hz are commonly detected in the HS and HIMS \citep{Ingram2019}. Although several models have been proposed to explain type-C QPOs \citep{2009MNRAS.397L.101I,2021NatAs...5...94M,2022NatAs...6..577M}, their physical origin remains a subject of ongoing debate.
%
%
By examining the energy-dependent properties of QPOs, we can gain a deeper understanding of their physical origin and the accretion flow geometry responsible for their production \citep[e.g.,][]{2022MNRAS.515.2099B,2022MNRAS.512.2686Z,2023ApJ...948..116M}.
The QPO fractional rms typically increases with energy initially and then turns to flat at around 10 keV \citep{2004ApJ...615..416R,2013MNRAS.433..412L,2016ApJ...833...27Y}. \hxmt\ observations have confirmed that the QPO fractional rms remains more or less constant up to 200 keV \citep{2018ApJ...866..122H,2023ApJ...948..116M}.
The behavior of the QPO lags is complex and not consistently patterned, particularly for the type-C QPOs \citep[e.g.,][]{2021NatAs...5...94M,2024MNRAS.527.9405M}.

\target\ is a new bright X-ray transient discovered on 2023 August 24 by MAXI/GSC \citep{2023ATel16205....1N,2023ATel16206....1N}, with a peak flux of exceeding 7 Crab in the MAXI 2--20 keV energy band. Subsequent multiwavelength observations have identified the source as a BH-LMXB \citep{2023ATel16208....1C,2023ATel16209....1W,2023ATel16210....1L,2023ATel16211....1M,2023ATel16217....1S}. 
%
\citet{2024ApJ...960L..17P} performed a spectral analysis using simultaneous \hxmt, {\it NICER}, and {\it NuSTAR} observations when the source was in the HIMS. They revealed the presence of an additional hard X-ray component that dominates the energy spectrum above 50 keV, a characteristic rarely observed in BH-LMXBs. 
Prominent type-C low-frequency QPOs with frequencies ranging from $\sim$0.1 to $\sim$8 Hz were detected in \target\ based on a timing analysis of \hxmt\ observations \citep{2024MNRAS.tmp..851Y,2024arXiv240509772Z}.
\citet{2024arXiv240417160N} reported the first detection of type-C QPOs up to 80--100 keV using {\it AstroSat/LAXPC} data.
%
%
In this Letter, we investigate the connection between the additional high-energy spectral component and the timing characteristics, and discuss the potential physical origin of this component. 

\section{Observations and Data Reduction}

\hxmt\ is China's first X-ray astronomy satellite launched on 2017 June 15 \citep{2020SCPMA..6349502Z}.
It carries three slat-collimated instruments: the High Energy X-ray telescope (HE, 20--250 keV, \citealt{2020SCPMA..6349503L}), the Medium Energy X-ray telescope (ME, 5--30 keV, \citealt{2020SCPMA..6349504C}), and the Low Energy X-ray telescope (LE, 1--15 keV, \citealt{2020SCPMA..6349505C}).

\begin{figure*}
\epsscale{0.5}
\centering
	\includegraphics[width=0.98\textwidth]{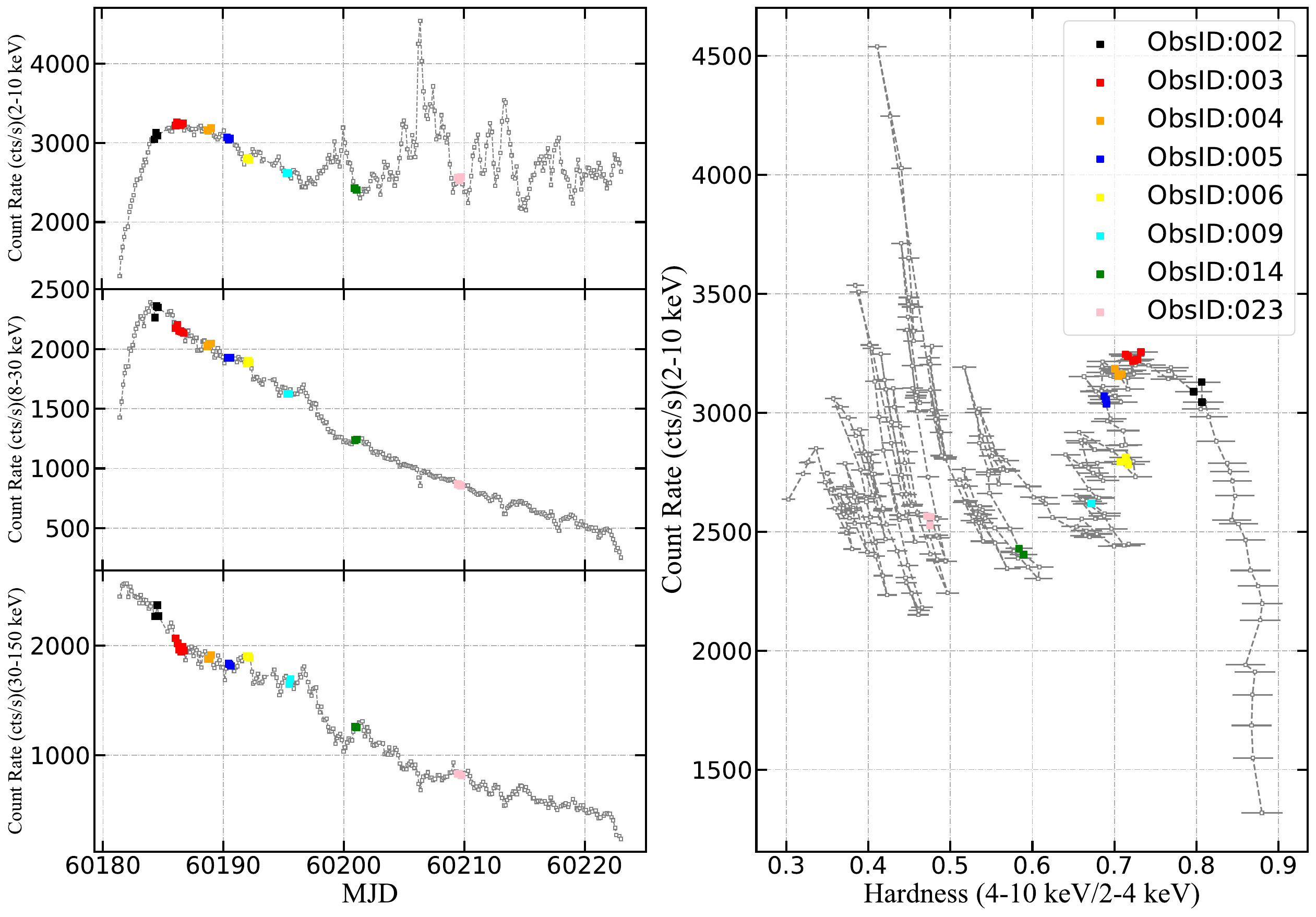}
    \caption{The LE (2--10 keV), ME (8--30 keV), HE (30--150 keV) light curves (left three panels) and the hardness intensity diagram (right panel) of the outburst of \target\ observed by \hxmt. The hardness is defined as the ratio between the 2--4 keV and 4--10 keV count rates. Each data point corresponds to an exposure ID. The different color points mark the observations we analyse in this paper.} 
    \label{lc_HID}
\end{figure*}

\hxmt\ has extensively monitored the outburst of \target\ between 2023 August 25 and 2023 October 4, corresponding to ObsIDs P0614338001--P0614338035. Note that each \hxmt\ observation (ObsID) is split into multiple segments (named ExpID).  In Figure~\ref{lc_HID}, we show the LE(2--10 keV), ME(8--30 keV), HE(30--150 keV) light curve and the hardness intensity diagram (HID) of the outburst. As we can see, the 2--10 keV count rate increases rapidly from below 1500 to above 3000 ${\rm cts}\ {\rm s}^{-1}$, reaching its peak at around MJD 60186. Subsequently, the source flux steadily decreases until MJD 60197, followed by 5 main flares, each lasting several days. As for the ME and HE light curves, no frequent flares along the outburst were found.
The evolution of the source along the HID indicates that the source was in the HS and HIMS during this period (see also \citealt{2024MNRAS.tmp..851Y}).

The data are extracted from all three instruments using the \hxmt\ Data Analysis software (HXMTDAS) v2.06\footnote{The data analysis software is available from \url{http://hxmten.ihep.ac.cn/software.jhtml}.}, and filtered with the following standard criteria:  
(1) pointing offset angle less than $0.04^{\circ}$;  
(2) Earth elevation angle larger than $10^{\circ}$;  
(3) the value of the geomagnetic cutoff rigidity larger than 8 GV;  
(4) at least 300 s before and after the South Atlantic Anomaly passage.
To avoid possible contamination from the bright Earth and nearby sources, we only use data from the small field of view (FoV) detectors \citep{2018ApJ...864L..30C,2022ApJ...937...33Y}.  The energy bands adopted for spectral analysis are 2--10 keV for LE, 8--28 keV for ME, and 28--120 keV for HE.

\begin{deluxetable*}{cccccc}[ht]
\tablecaption{The log of the \hxmt\ observations used in this work for the energy-dependent analysis. The hardness is defined as the ratio of the count rate between the 2.0--4.0 keV and the 4.0--10.0 keV bands. The QPO frequency is extracted from the LE 2--10 keV PDS. We used the last three digits of each ObsID in the text and figures for clarity. \label{log}}
\tablehead{
\colhead{ObsID} & \colhead{ExpID} & \colhead{Start time (MJD)} & \colhead{2--10 keV rate (cts ${\rm s}^{-1}$)} & \colhead{Hardness} & \colhead{QPO Frequency (Hz)}
}
\startdata
  P0614338002 & P0614338002010 & 60184.26 & $3088\pm1$ & $0.80\pm0.02$ & $0.71\pm0.01$ \\
  {}          & P0614338002011 &  &  &  & \\
  {}          & P0614338002012 &  &  &  & \\
  \hline
  P0614338003 & P0614338003006  & 60185.99  & $3234\pm1$ & $0.72\pm0.02$ & $1.12\pm0.01$\\
  {}          & P0614338003007  &   &  &  & \\
  {}          & P0614338003008  &   &  &  & \\
  {}          & P0614338003009  &   &  &  & \\
  {}          & P0614338003010  &   &  &  & \\
  {}          & P0614338003011  &   &  &  & \\
  \hline
  P0614338004 & P0614338004012 &  60188.63  & $3168\pm1$  & $0.71\pm0.02$ & $1.24\pm0.01$\\
  {}          & P0614338004013 &    &  &  & \\
  {}          & P0614338004014 &    &  &  & \\
  \hline
  P0614338005 & P0614338005010 & 60190.28    & $3054\pm1$ & $0.69\pm0.02$  & $1.30\pm0.01$\\
  {}          & P0614338005011 &    &  &  & \\
  {}          & P0614338005012 &    &  &  & \\
  \hline
  P0614338006 & P0614338006007 & 60191.87    & $2797\pm1$  & $0.71\pm0.02$ & $1.12\pm0.01$\\
  {}          & P0614338006008 &    &  &  & \\
  {}          & P0614338006009 &    &  &  & \\
  \hline
  P0614338009 & P0614338009002 &  60195.23   & $2619\pm1$ & $0.67\pm0.02$ & $1.40\pm0.01$\\
  {}          & P0614338009003 &    &  &  & \\
  \hline
  P0614338014 & P0614338014007 &  60200.85   & $2417\pm1$ & $0.58\pm0.02$ & $2.05\pm0.01$\\
  {}          & P0614338014008 &    &  &  & \\
  \hline
  P0614338023 & P0614338023003 & 60209.38    & $2552\pm1$ & $0.48\pm0.01$ & $3.85\pm0.01$\\
  {}          & P0614338023004 &    &  &  & \\
  {}          & P0614338023005 &    &  &  & \\
  \hline
  \hline
\enddata
\end{deluxetable*}

\section{Analysis and Results} 

\begin{figure}[ht!]
\centering
	\includegraphics[width=0.48\textwidth]{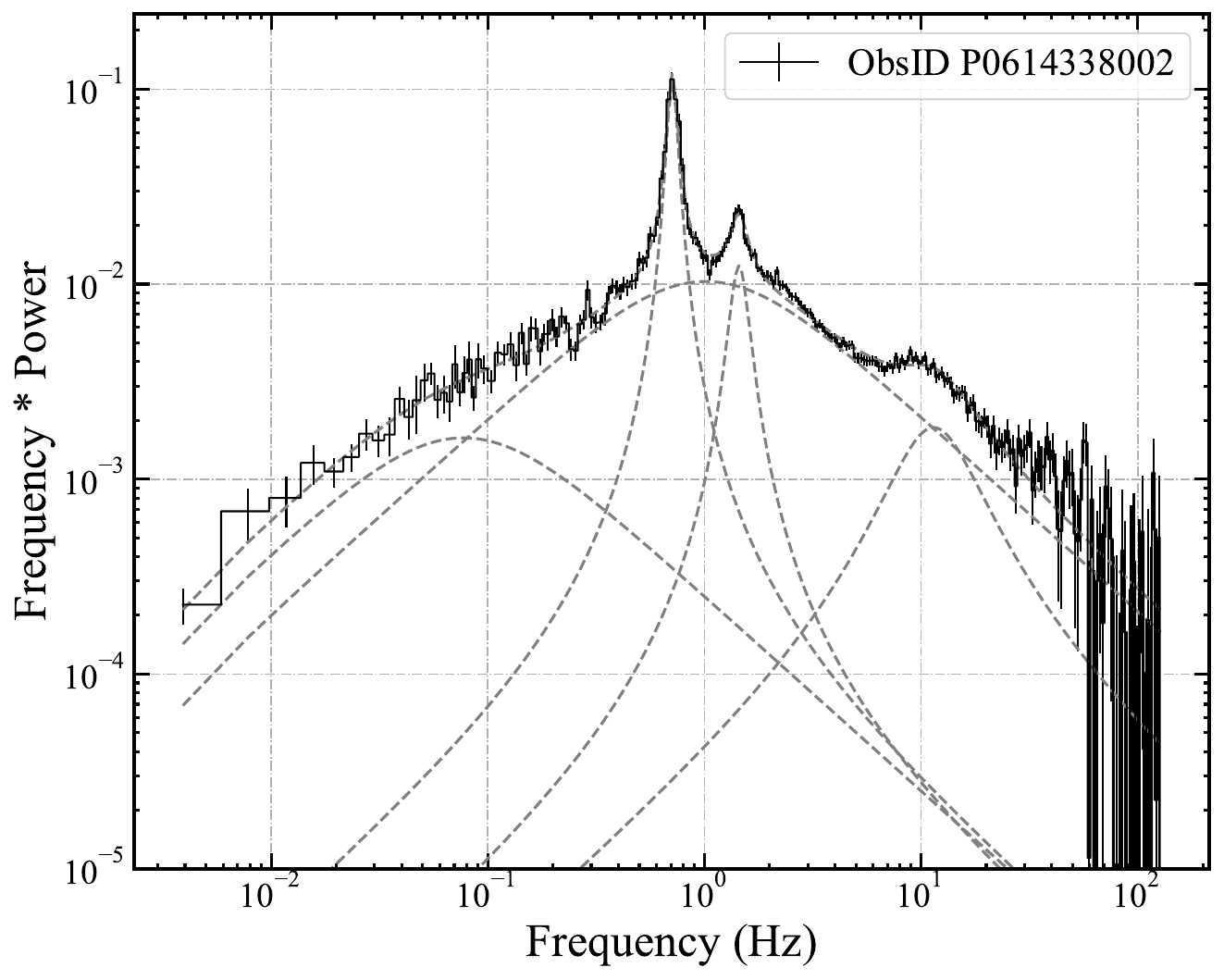}
    \caption{A representative power density spectrum calculated in the 2--10 keV energy band for ObsID P0614338002. The power density spectrum was fitted with a multi-Lorentzian model. 
    %
    }
    \label{PDS}
\end{figure}

We first checked the power density spectrum (PDS) in different energy bands for each ExpID. We used a 128-s long interval and a 1/256-s time resolution. The PDS is applied to Miyamoto normalization and the Possion noise is subtracted \citep{1991ApJ...383..784M}. We fitted the PDS with a multi-Lorentzian model \citep{2002ApJ...572..392B} and obtained the QPO parameters. To study the energy-dependent properties of the QPO, we selected eight observations with relatively long exposure times covering both the normal decay phase and flares. For each observation, we combined ExpIDs with similar PDS shapes to enhance the statistics. The observations used for the energy-dependent analysis are listed in Table~\ref{log} and marked in Figure~\ref{lc_HID} with different colors. We then extracted PDS in multiple energy bands for each observation and fitted them with the multi-Lorentzian model. A representative PDS in the 2--10 keV energy band is shown in Figure~\ref{PDS}. A strong type-C QPO with its second harmonic is seen in the PDS.


\begin{figure*}[!ht]
\epsscale{0.5}
\centering
	\includegraphics[width=0.98\textwidth]{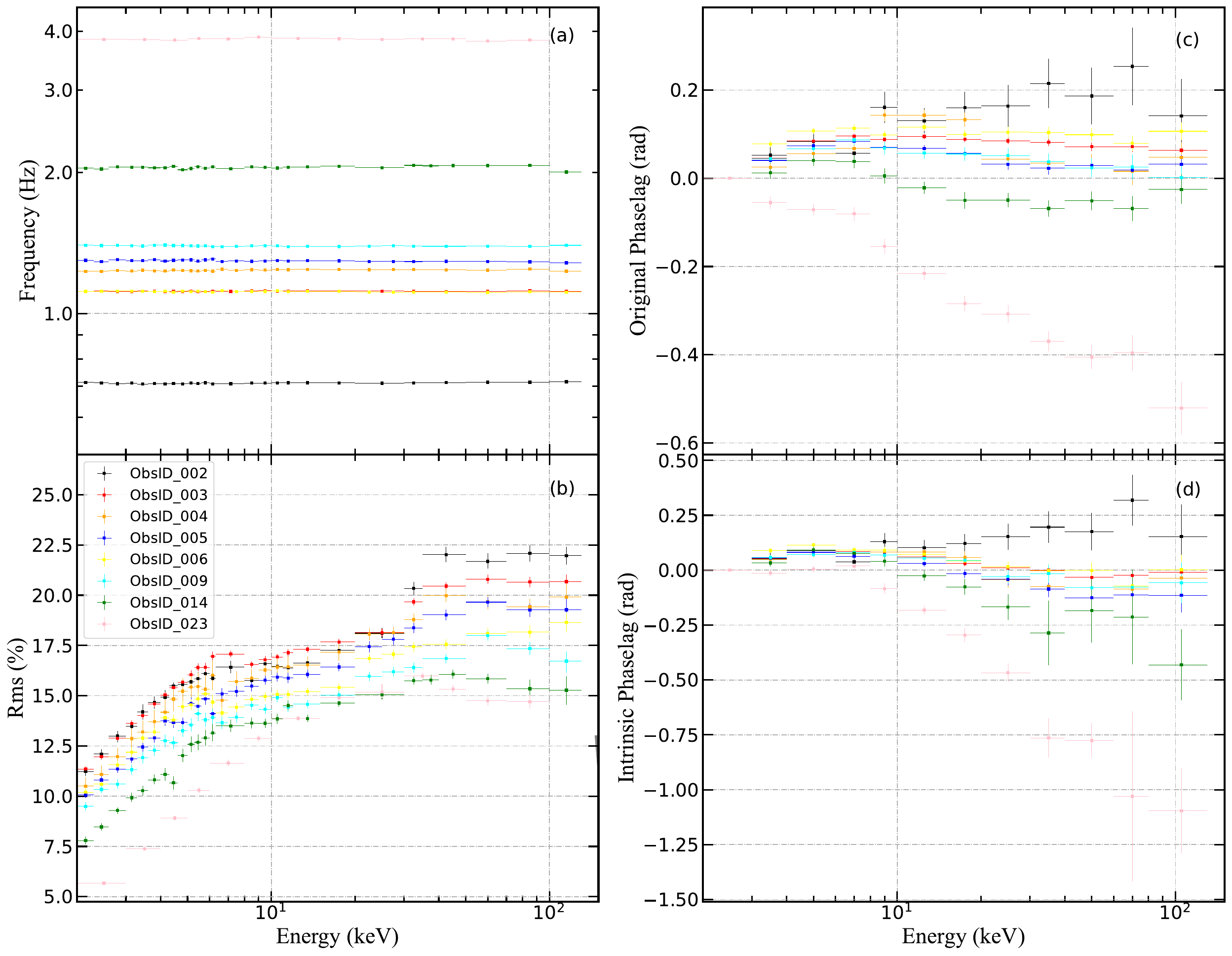}
    \caption{QPO centroid frequency (panel a), fractional rms (panel b), original (panel c) and intrinsic phase lags (panel d) as a function of photon energy for different observations of \target.}
    \label{rms_spec}
\end{figure*}

In the panel (a) and (b) of Figure~\ref{rms_spec}, we show the centroid frequency and fractional rms of the QPO as a function of photon energy for all observations, respectively.
The energy bands we used are LE (1--1.3, 1.3--1.6, 1.6--2.0, 2.0--2.3, 2.3--2.6, 3.6--3.0, 3.0--3.3, 3.3--3.6, 3.6--4.0, 4.0--4.3, 4.3--4.6, 4.6--5.0, 5.0--5.3, 5.3--5.6, 5.6--6.0, 6.0--6.3 keV), ME (6.3--8.0, 8--9, 9--10, 10--11, 11--12, 12--15, 15--20, 20--30 keV), HE (30--35, 35--50, 50--70, 70--100, 100--130 keV).
We find that the QPO frequency does not change with energy for all observations.
The QPO fractional rms reflects the ratio of the variability amplitude of X-ray flux to its mean intensity. 
The QPO rms spectra of many BH-LMXBs observed with Insight-HXMT in a broad energy band (2--200 keV) show a common feature: the QPO fractional rms first increases with photon energy from 2 keV to about 7--10 keV and then start to flatten up to 100 keV and even higher energies \citep{2018ApJ...866..122H,2020JHEAp..25...29K,2021ApJ...919...92B,2022MNRAS.512.2686Z,2023ApJ...948..116M}. 
At energies below 20 keV \target\ shows a similar behavior. However, at energies above 20-40 keV, \target\ exhibits a distinct behavior: its QPO fractional rms above 20-40 keV is significantly higher than that below 20 keV and remains almost constant at high energies up to above 100 keV, which we call high energy (HE)-rms excess. This is the first time such a HE-rms excess is found in a BH-LMXB.
The significance of the HE-rms excess diminishes over time as the source softens.

Following \citet{2021NatAs...5...94M}, we also investigated the energy-dependent QPO phase lag and calculated the QPO original and intrinsic phase lags in different energy bands, with reference to the 2--3 keV band. The original phase lags were calculated by averaging the phase lag over the QPO frequency range $\nu_{0} \pm {\rm FWHM}/2$ in the lag-frequency spectra. The average value of data points below the QPO frequency in the lag–frequency spectra is considered as the phase-lag continuum. The intrinsic QPO lags were determined by subtracting the phase-lag continuum from the original phase lag at the QPO frequency. In the panel (c) and (d) of Figure~\ref{rms_spec}, we show the QPO original and intrinsic phase lags as a function of photon energy, respectively. Notably, the shape of the QPO phase-lag spectrum follows a nearly log-linear dependence on energy. No significant differences are observed between energies below and above 30 keV.   

\begin{figure}[!ht]
\epsscale{0.5}
\centering
	\includegraphics[width=0.48\textwidth]{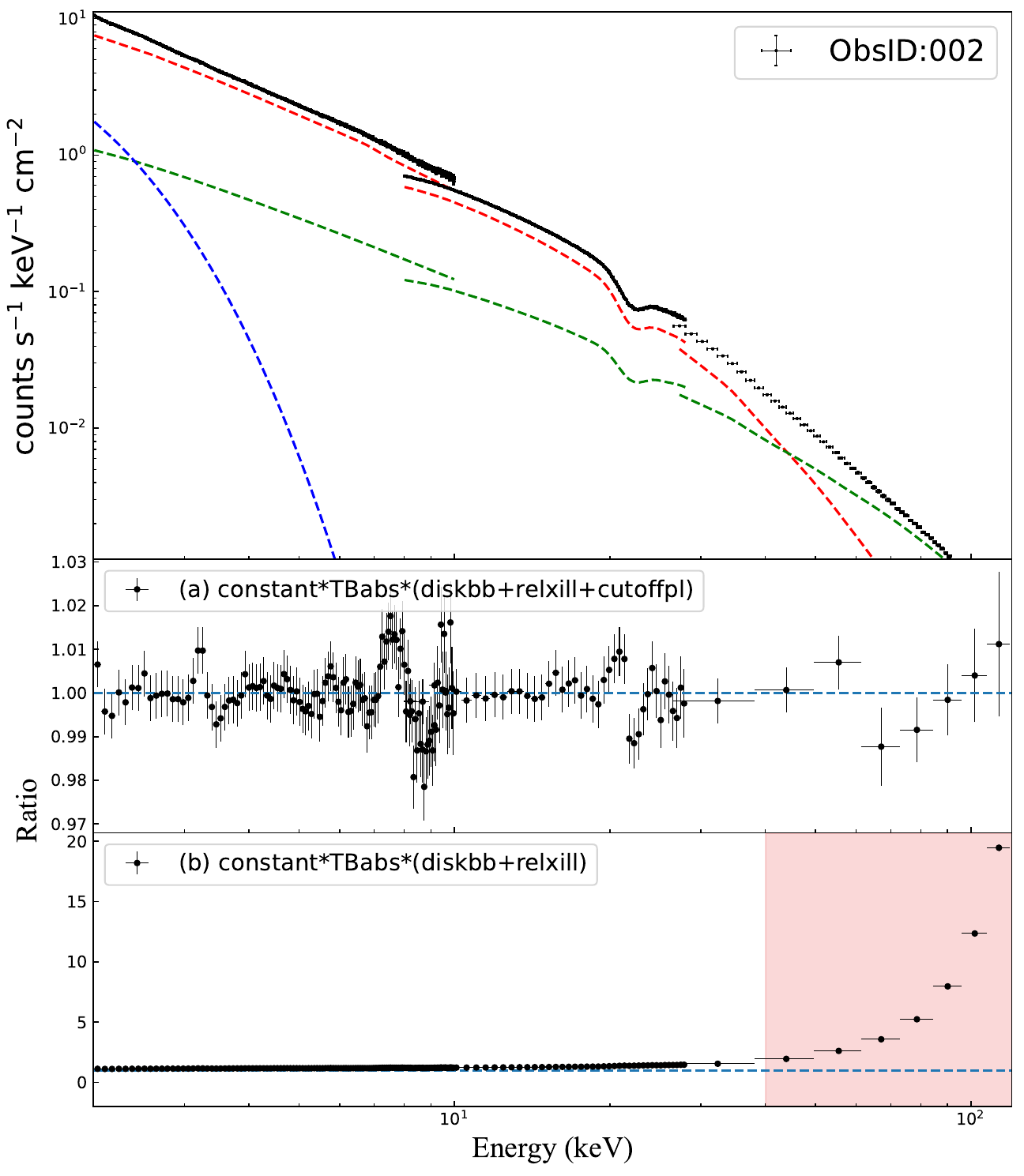}
    \caption{Top and middle panels: Folded spectra and residuals fitted with the model {\tt constant*TBabs*(diskbb+relxill+cutoffpl)} for ObsID P0614338002. The blue, red, green lines mark the {\tt diskbb}, {\tt relxill} and {\tt cutoffpl} component separately.  Bottom panel: the spectral residuals using the same model but setting the normalization of cutoffpl to zero. The red region marks the 40--120 keV energy band with the HE-rms excess.}
    \label{spec_residual}
\end{figure}

Based on a spectral analysis using simultaneous \hxmt, {\it NICER} and {\it NuSTAR} observations, \citet{2024ApJ...960L..17P} found an extra hard component in addition to the standard thermal Comptonization component in the spectrum of \target. 
For each observation we used, we first tried to fit the spectra with one single hard component plus a disk component modified by absorption, e.g., {\tt constant*TBabs*(diskbb+cutoffpl+gaussian)}, {\tt constant*TBabs*(thcomp*diskbb+gaussian)}. The {\tt constant} reflects the relative calibration between LE, ME, HE instruments, and was fixed to 1 for LE. The {\tt TBabs} model is used to account for the interstellar absorption with abundances from \citet{2000ApJ...542..914W}. The {\tt diskbb} model is a multi-temperature blackbody spectrum to explain the emission from the accretion disk \citep{1984PASJ...36..741M}. The {\tt thcomp} model is a novel thermal Comptonization convolution model to describe the hard power-law component, which is presented as a replacement for the {\tt nthcomp} model \citep{2020MNRAS.492.5234Z}. We find that both models give unacceptable fitting results, exhibiting large residuals at high energy bands and the iron line at 6--7 keV.
Thus, in order to compensate the high-energy residuals and explain the reflection features, we adopted the model {\tt constant*TBabs*(diskbb+relxill+cutoffpl)}, where the novel {\tt relxill} accounts for the standard thermal Comptonization component with reflection \citep{2014ApJ...782...76G} and the {\tt cutoffpl} corresponds to the additional hard component seen in the spectra.
During fittings, we fixed the neutral absorption parameter, $N_{\rm H}$, at $0.3\times10^{22} \ {\rm cm}^{-2}$ following \citet{2024arXiv240603834L}. The spin and inclination angle were fixed at the values obtained by \citet{2024ApJ...960L..17P}. Additionally, the iron abundance, $A_{\rm Fe}$, was linked between all the spectra. We added a systematic error of 0.5 percent for the spectra.
In the top and middle panels of Figure~\ref{spec_residual}, we show the unfolded spectra and residuals for ObsID P0614338002 with the best-fitting model. In the bottom panel, we show the spectral residuals by setting the normalization of {\tt cutoffpl} to zero to illustrate the additional hard component. We can see that the additional hard component mainly dominates the energy bands above 30--40 keV. 
This model gives acceptable spectral fits for all observations. The best-fitting results are listed in Table~\ref{spec_reflection}. 
We found that the photon index $\Gamma$ of the standard Comptonization component increases from $1.47^{+0.01}_{-0.04}$ to $2.17\pm0.01$ with time. As the source softens, the inner disk temperature gradually increases from around $0.46\pm0.01$ keV to $0.89\pm0.01$ keV.
The inner radius of the accretion disk is very close to the innermost stable circular orbit with a slight decreasing trend from $R_{\rm in}=4.0\pm0.4~R_{\rm g}$ to $R_{\rm in}=2.8\pm0.2~R_{\rm g}$.
Regarding the additional {\tt cutoffpl} component, the cut-off energy $E_{\rm cut}$ is significantly higher than that of {\tt relxill}, increasing from $\sim$50 keV to above 100 keV with time. Additionally, the flux of {\tt cutoffpl} decreases gradually, particularly evident in the last two observation.

%

\begin{figure*}[!ht]
\epsscale{0.5}
\centering
	\includegraphics[width=0.98\textwidth]{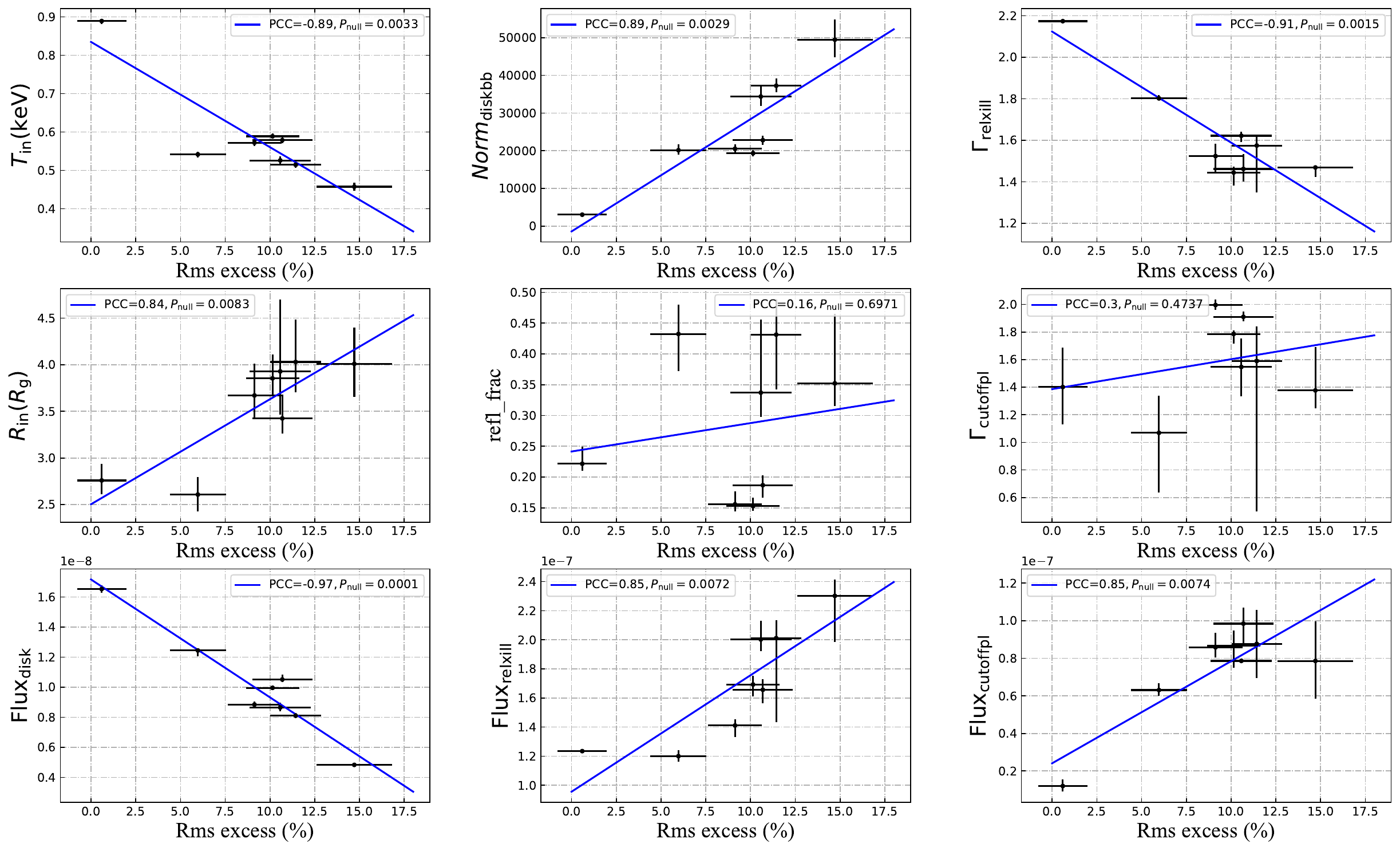}
    \caption{The relation between the value of the HE-rms excess and the spectral parameters. The spectra were fitted with the model  {\tt constant*TBabs*(diskbb+relxill+cutoffpl)}. The spectral parameters plotted are the inner disk temperature $T_{\rm in}$, {\tt diskbb} normalization $Norm_{\rm diskbb}$, the photon index of {\tt relxill} $\Gamma_{\rm relxill}$, inner disk radius $R_{\rm in}$, reflection fraction refl\_frac, the photon index of {\tt cutoffpl} $\Gamma_{\rm cutoffpl}$, and the fluxes of the {\tt diskbb}, {\tt relxill}, {\tt cutoffpl} components. The unabsorbed fluxes of the different components are obtained in the 2--120 keV band using {\tt cflux}. The solid line is the best-ﬁtting straight line to the data. The Pearson Correlation Coefficient (PCC) and  the null hypothesis probability are labelled in the legend.}
    \label{rms_parameter}
\end{figure*}

Given the simultaneous presence of the HE-rms excess and the additional hard spectral component, it is imperative to investigate the connection between the HE-rms excess and the main spectral parameters. To accurately quantify the HE-rms excess, we used a phenomenological function $\sigma(E)=\sigma_0/(1+e^{-k(E-E_0)})$ to fit the 2--20 keV rms-energy relation \citep{2020JHEAp..25...29K}, which exhibits a rising trend followed by a plateau, similar to that seen in many other BH-LMXBs. The maximum value $\sigma_0$ corresponds to the plateau value around 8--20 keV. Subsequently, we calculated the average QPO rms values above 40 keV. The HE-rms excess was then calculated by subtracting in quadrature the plateau value from the average QPO rms above 40 keV (red shaded region in Figure 4). In Figure~\ref{rms_parameter}, we show the relation between the value of the HE-rms excess and the main spectral parameters obtained by fitting the spectra with the model {\tt constant*TBabs*(diskbb+relxill+cutoffpl)}. 
We find that the HE-rms excess follows a tight linear correlation with the inner disk temperature $T_{\rm in}$, as well as the normalization of {\tt diskbb} and the flux of the disk component, with a significance exceeding 3$\sigma$. Furthermore, we find that the HE-rms excess is negatively correlated with the photon index of the standard Comptonization component and moderately correlated with its flux, indicating that the HE-rms excess diminishes as the spectrum softens. We also find that the HE-rms excess exhibits moderate positive correlation with the flux of the additional {\tt cutoffpl} component, with a null hypothesis probability of $\sim$ 0.007.
%
%
%

\begin{deluxetable*}{cccccccccc}
\tablecaption{Best-fitting results using the model {$\tt constant\ast \tt TBabs\ast (\tt diskbb + \tt relxill + \tt cutoffpl)$}. The fluxes are calculated in the 2--120 keV band, and presented in the unit of $10^{-8} \ {\rm erg} \ {\rm s}^{-1} \ {\rm cm}^{-2}$.   \label{spec_reflection}}
\tabletypesize{\scriptsize}
\tablehead{
\colhead{Component} & \colhead{Parameter} & \colhead{002} & \colhead{003} & \colhead{004} & \colhead{005} & \colhead{006} & \colhead{009} & \colhead{014} & \colhead{023}
}
\startdata
constant  &   LE     & \multicolumn{8}{c}{1 (fixed)}\\
{}        &   ME     & $0.95\pm0.01$ & $0.95\pm0.01$ & $0.94\pm0.01$ & $0.95\pm0.01$ & $0.95\pm0.01$ & $0.95\pm0.01$ & $0.96\pm0.01$ & $0.98\pm0.01$ \\
{}        &   HE     & $0.91\pm0.01$ & $0.92\pm0.01$ & $0.90\pm0.01$ & $0.91\pm0.01$ & $0.91\pm0.01$ & $0.92\pm0.01$ & $0.93\pm0.01$ & $0.97\pm0.01$ \\
\hline
TBabs     &   $N_{\rm H}(\times 10^{22} {\rm cm}^{-2})$ & \multicolumn{8}{c}{0.3 (fixed)} \\
\hline
Diskbb    &   $T_{\rm in} (\rm keV)$ & $0.46\pm0.01$ & $0.51\pm0.01$ & $0.53\pm0.01$ & $0.58\pm0.01$ &$0.59\pm0.01$ & $0.57\pm0.01$ & $0.54\pm0.01$ & $0.89\pm0.01$ \\
{}        &   norm ($10^4$)   & $4.95^{+0.54}_{-0.47}$ & $3.72^{+0.18}_{-0.17}$ & $3.44^{+0.28}_{-0.25}$ & $2.28\pm0.12$ & $1.93\pm0.08$ & $2.05\pm0.12$ & $2.01^{+0.15}_{-0.12}$ & $0.30\pm{0.01}$\\
\hline
Relxill   & Index1  & \multicolumn{8}{c}{3.0 (fixed)} \\
{}        & Index2  & \multicolumn{8}{c}{3.0 (fixed)}\\
{}        & Rbr     & \multicolumn{8}{c}{15 (fixed)}\\
{}        & a       & \multicolumn{8}{c}{0.98 (fixed)}\\
{}        & Incl    & \multicolumn{8}{c}{40 (fixed)}\\
{}        & Rin     & $4.0\pm0.4$ & $4.0^{+0.5}_{-0.3}$ & $3.9^{+0.8}_{-0.5}$ & $3.4^{+0.3}_{-0.2}$ & $3.9^{+0.3}_{-0.2}$ & $3.7^{+0.3}_{-0.2}$ & $2.6\pm0.2$ & $2.8\pm0.2$\\
{}        & Rout    & \multicolumn{8}{c}{400 (fixed)}\\
{}        & z       & \multicolumn{8}{c}{0 (fixed)}\\
{}        & gamma   & $1.47^{+0.01}_{-0.04}$ & $1.57^{+0.03}_{-0.22}$ & $1.62\pm0.03$ & $1.46\pm0.07$ & $1.44^{+0.03}_{-0.06}$ & $1.52^{+0.06}_{-0.08}$ & $1.80\pm0.01$ & $2.17\pm0.01$\\
{}        & logxi   & $1.91\pm0.10$ & $1.78\pm0.01$ & $1.91^{+0.13}_{-0.17}$ & $2.96^{+0.16}_{-0.08}$ & $3.00^{+0.07}_{-0.03}$ & $2.98\pm0.09$ & $3.94^{+0.08}_{-0.04}$ & $3.84\pm0.09$\\ 
{}        & Afe     & \multicolumn{8}{c}{1.17 (linked)}\\
{}        & Ecut (keV)    & $16\pm1$ & $16\pm1$ & $17\pm1$ & $17\pm1$ & $18\pm1$ & $20\pm1$ & $35\pm1$ & $47^{+3}_{-2}$\\
{}        & refl\_frac & $0.35^{+0.11}_{-0.04}$ & $0.43^{+0.05}_{-0.09}$ & $0.34^{+0.12}_{-0.04}$ & $0.19\pm0.02$ & $0.15\pm0.01$ & $0.16\pm0.02$ & $0.43^{+0.04}_{-0.06}$ & $0.22\pm0.02$\\
{}        & norm    & $0.38\pm0.02$ & $0.36^{+0.07}_{-0.02}$ & $0.38^{+0.07}_{-0.05}$ & $0.27\pm0.05$ & $0.27\pm0.04$ & $0.23\pm0.04$ & $0.28\pm0.01$ & $0.43\pm0.01$\\
\hline
cutoffpl &  $\Gamma$ & $1.4^{+0.3}_{-0.1}$ & $1.6^{+0.3}_{-1.0}$ & $1.6\pm0.2$ & $1.91\pm0.04$ & $1.79^{+0.03}_{-0.07}$ & $2.00\pm0.04$ & $1.1^{+0.3}_{-0.4}$ & $1.4\pm0.3$ \\
{}        &  $E_{\rm cut} (\rm keV)$ & $50^{+3}_{-6}$ & $57^{+5}_{-3}$ & $57^{+8}_{-28}$ & $87\pm7$ & $79^{+3}_{-6}$ & $117^{+19}_{-15}$ & 300(fixed) & 300(fixed)  \\
{}        &  norm & $4\pm2$ & $7\pm4$ & $6^{+10}_{-4}$ & $16^{+6}_{-3}$ & $10\pm4$ & $16^{+6}_{-3}$ & $0.06^{+0.11}_{-0.05}$ & $0.20^{+0.67}_{-0.09}$ \\
\hline
Flux  & Diskbb & $0.48\pm0.01$ & $0.81\pm0.02$ & $0.87\pm0.03$ & $1.05^{+0.03}_{-0.02}$ & $1.00\pm0.01$ & $0.88\pm0.02$ & $1.24^{+0.01}_{-0.04}$ & $1.65\pm0.02$\\
Flux  & Relxill & $23^{+1}_{-3}$ & $20^{+1}_{-6}$ & $20\pm1$ & $17\pm1$ & $17\pm1$ & $14\pm1$ & $12\pm1$ & $12\pm1$ \\
Flux  & Cutoffpl & $8\pm2$ & $9\pm2$ & $7.9\pm0.1$ & $10\pm1$ & $8.7^{+0.8}_{-1.2}$ & $8.6^{+0.8}_{-0.5}$ & $6.3^{+0.4}_{-0.3}$ & $12.0\pm0.3$ \\
\hline
{}        & $\chi^2_{red}/d.o.f$ & $1237.26/1337$ & $1282.54/1337$ & $1231.47/1336$ & $1300.38/1337$ & $1490.21/1337$ & $1300.90/1337$ & $1370.07/1316$ & $1234.97/1338$ \\
\hline
\hline
\enddata
\end{deluxetable*}

\section{Discussion and Conclusion} 
In this Letter, we investigated the energy-dependent properties of the type-C QPOs observed in the newly discovered BH-LMXB \target\ using \hxmt\ data. 
A significant HE-rms excess at energies above 20--40 keV is detected for the first time in BH-LMXBs, thanks to the large effective area of \hxmt\ in the high-energy band. This deviation from the typical flat QPO rms spectrum above 10 keV, commonly observed in many other BH-LMXBs \citep{2004A&A...426..587C,2004ApJ...615..416R,2012Ap&SS.337..137Y,2013MNRAS.434...59Y,2016ApJ...833...27Y,2018ApJ...866..122H,2021NatAs...5...94M}, suggests a unique behavior of \target.
We note that  \citet{2024MNRAS.tmp..851Y} and \citet{2024arXiv240509772Z} also studied the energy-dependent QPO fractional rms using a few \hxmt\ observations of \target. However, in their work, they used a low energy resolution in producing the QPO rms spectra so that the characteristics of the HE-rms excess is not clearly seen.
\citet{2024arXiv240417160N} studied the evolution of the QPO fractional rms with photon energy using {\it AstroSat} observations. However, they found that the QPO rms decreases from $\sim$13 percent at 20--40 keV to $\sim$2.5 percent at 80--100 keV, which is inconsistent with the trend we find. We propose that the difference may be due to that the background is not considered in their calculation.
%

In the HS and HIMS, the spectra of BH-LMXBs are usually dominated by a hard power-law component, which is believed to originate from thermal Comptonization of soft disk photons by hot electrons in the corona/jet base.
In addition to the standard hard power-law component, a high-energy hard tail is sometimes observed in the spectra of BH-LMXBs \citep[e.g.,][]{2006A&A...446..591C,2021ApJ...909L...9Z,2023A&A...669A..65C}. The physical origin of the hard tail is still under debate. 
\citet{2024ApJ...960L..17P} reported the detection of a high-energy hard tail in the spectra of \target.  Our spectral analysis confirms the presence of this component and reveals that it gradually weakens as the spectrum softens.
%
%
The observed QPO HE-rms excess and the additional high-energy spectral component are found in similar energy bands, indicating a potential physical connection between them. We did find a moderate positive correlation between the value of the HE-rms excess and the flux of the additional hard component. This suggests that the QPO HE-rms excess is probably produced in the additional high-energy component. Additionally, as shown in Figure~\ref{rms_parameter}, the HE-rms excess also strongly depends on the variation of the disk parameters. As the disk emission increases, both the additional hard component and the HE-rms excess become weaker. This suggests that the variation of accretion disk might be the primary and intrinsic driving factor influencing the strength of the HE-rms excess.

Phase lag between QPO signals in different energy bands serves a powerful tool for  constraining the geometry of the accretion flow that generates the QPO. We have studied the energy-dependent phase lags of the QPO in \target. However, we did not observe any clear features corresponding to the additional hard component seen at high energies. The QPO phase lags exhibit a smooth evolution across all energy bands, implying a similarity in the geometry between the emitting region responsible for the additional hard component and the standard hard power-law component.
In addition, it is worth noting that the QPO frequency does not change with photon energy (see Figure~\ref{rms_spec} and \citealt{2024MNRAS.tmp..851Y}). This means that the additional hard component is modulated at the same frequency as the standard hard power-law component. Therefore, it is likely that the two different hard components originate from different radiation mechanisms within the same region. 
%
%
Two scenarios have been proposed to explain the additional high-energy tail seen in BHXBs. The first scenario considers the Synchrotron spectrum from the jet base \citep{2005ApJ...635.1203M,2011Sci...332..438L,2014ApJ...789...26J,2015ApJ...807...17R,2021MNRAS.500.2112K}. The high polarization degree of the high-energy tail detected in Cyg X-1 and MAXI J1348--630 is consistent with the emission from the compact jet of these sources \citep{2012ApJ...761...27J,2023A&A...669A..65C}. 
The alternative scenario suggests that the high-energy tail arises from a hybrid thermal/non-thermal Comptonization in a plasma with a non-thermal electron distribution. The hybrid Comptonization model, e.g. {\tt eqpair}, has been successfully used to fit the  high-energy tail in a few sources \citep{2006A&A...446..591C,2022ApJ...928...11Z,2023A&A...669A..65C}. We find correlations between the disk parameters and the HE-rms excess (or the additional hard component). This result supports the hybrid Comptonization model as the soft photons from accretion disk serve as natural seed photons for Comptonization.
In both cases, the jet base is a potential region that produces the additional high-energy component. Especially, recent polarization observation by INTEGRAL for \target\ reveals that the polarization fraction reaches up to 50\% in the 200--400 keV, which supports the optically thin synchrotron emission from jet \citep{2024arXiv240410705B}.
To date, there is some evidence supporting that type-C QPOs may originate from the Lense-Thirring precession of a small-scale jet \citep[e.g.,][]{2021NatAs...5...94M}. In this model, the QPO HE-rms excess can be easily explained as the additional high-energy component produced in the jet is modulated by the precession and contributes to the QPO rms in addition to the standard thermal Compotonization component.
On the other hand, the IXPE polarization observations of a few BH-LMXBs during their HS support a horizontal elongated corona configuration \citep{2022Sci...378..650K}. \citet{2023ApJ...958L..16V} report the first detection of the X-ray polarization of \target\ for the observation carried out in HIMS, which happened close to the observation P0614338009 we used in this paper. They found a time- and energy-averaged polarization degree of $4.1\%\pm 0.2\%$ and a polarization angle of $2^{\circ}.2\pm1^{\circ}.3$. Combined with the submillimeter band observation, these results indicate that the corona is elongated orthogonal to the jet. Similar accretion geometry was also proposed in hard state by \citet{2023arXiv231105497I,2024arXiv240419601P}. As the outburst evolves, \citet{2024arXiv240304689S} found that the polarization degree decreases dramatically to $\leq 1\%$ in the soft state, which suggested that the polarization is mainly contributed by the up-scattered radiation in the hot corona. Meanwhile, based on a phase-resolved polarimetric analysis, \citet{2024ApJ...961L..42Z} found no modulations of polarization degree and polarization angle with QPO phase in the 2--8 keV band for the same observation. This result poses a challenge to the hot flow precession model for \target. This may be because the regions responsible for producing polarization and generating the QPO modulation are different. The QPO signal mainly arises from the precession of the jet, and the radiation from the jet contributes to a relatively low proportion of emission in the 2–8 keV range, leading to the polarization modulation being less significant with QPO phase.

\begin{acknowledgments}
We are grateful to the referee for the constructive comments to improve the manuscript. This work has made use of the data from the \hxmt\ mission, a project funded by China National Space Administration (CNSA) and the Chinese Academy of Sciences (CAS), and data and/or software provided by the High Energy Astrophysics Science Archive Research Center (HEASARC), a service of the Astrophysics Science Division at NASA/GSFC. This work is supported by the National Key RD Program of China (2021YFA0718500), International Partnership Program of Chinese Academy of Sciences under (Grant No. 113111KYSB20190020) and the National Natural Science Foundation of China (NSFC) under grants 12203052,12122306, 12333007.
\end{acknowledgments}
%

\vspace{5mm}





\bibliography{sample631}{}

\begin{thebibliography}{}
\expandafter\ifx\csname natexlab\endcsname\relax\def\natexlab#1{#1}\fi
\providecommand{\url}[1]{\href{#1}{#1}}
\providecommand{\dodoi}[1]{doi:~\href{http://doi.org/#1}{\nolinkurl{#1}}}
\providecommand{\doeprint}[1]{\href{http://ascl.net/#1}{\nolinkurl{http://ascl.net/#1}}}
\providecommand{\doarXiv}[1]{\href{https://arxiv.org/abs/#1}{\nolinkurl{https://arxiv.org/abs/#1}}}

\bibitem[{{Bellavita} {et~al.}(2022){Bellavita}, {Garc{\'\i}a}, {M{\'e}ndez},
  \& {Karpouzas}}]{2022MNRAS.515.2099B}
{Bellavita}, C., {Garc{\'\i}a}, F., {M{\'e}ndez}, M., \& {Karpouzas}, K. 2022,
  \mnras, 515, 2099, \dodoi{10.1093/mnras/stac1922}

\bibitem[{{Belloni} {et~al.}(2002){Belloni}, {Psaltis}, \& {van der
  Klis}}]{2002ApJ...572..392B}
{Belloni}, T., {Psaltis}, D., \& {van der Klis}, M. 2002, \apj, 572, 392,
  \dodoi{10.1086/340290}

\bibitem[{{Belloni} \& {Motta}(2016)}]{2016ASSL..440...61B}
{Belloni}, T.~M., \& {Motta}, S.~E. 2016, in Astrophysics and Space Science
  Library, Vol. 440, Astrophysics of Black Holes: From Fundamental Aspects to
  Latest Developments, ed. C.~{Bambi}, 61, \dodoi{10.1007/978-3-662-52859-4_2}

\bibitem[{{Belloni} {et~al.}(2011){Belloni}, {Motta}, \&
  {Mu{\~n}oz-Darias}}]{2011BASI...39..409B}
{Belloni}, T.~M., {Motta}, S.~E., \& {Mu{\~n}oz-Darias}, T. 2011, Bulletin of
  the Astronomical Society of India, 39, 409, \dodoi{10.48550/arXiv.1109.3388}

\bibitem[{{Bouchet} {et~al.}(2024){Bouchet}, {Rodriguez}, {Cangemi}, \&
  {Laurent}}]{2024arXiv240410705B}
{Bouchet}, T., {Rodriguez}, J., {Cangemi}, F., \& {Laurent}, P. 2024, arXiv
  e-prints, arXiv:2404.10705, \dodoi{10.48550/arXiv.2404.10705}

\bibitem[{{Bu} {et~al.}(2021){Bu}, {Zhang}, {Santangelo}, {Belloni}, {Zhang},
  {Qu}, {Tao}, {Huang}, {Ma}, {Li}, {Zhang}, {Chen}, {Cai}, {Cao}, {Chang},
  {Chen}, {Chen}, {Chen}, {Cui}, {Du}, {Gao}, {Gao}, {Ge}, {Gu}, {Guan}, {Guo},
  {Han}, {Huo}, {Jia}, {Jiang}, {Jin}, {Kong}, {Li}, {Li}, {Li}, {Li}, {Li},
  {Li}, {Li}, {Li}, {Li}, {Liang}, {Liao}, {Liu}, {Liu}, {Liu}, {Liu}, {Lu},
  {Lu}, {Luo}, {Luo}, {Ma}, {Meng}, {Nang}, {Nie}, {Ou}, {Sai}, {Song}, {Song},
  {Sun}, {Tan}, {Tuo}, {Wang}, {Wang}, {Wang}, {Wang}, {Wang}, {Wen}, {Wu},
  {Wu}, {Wu}, {Xiao}, {Xiao}, {Xiong}, {Xu}, {Yang}, {Yang}, {Yi}, {Yin},
  {You}, {Zhang}, {Zhang}, {Zhang}, {Zhang}, {Zhang}, {Zhang}, {Zhang},
  {Zhang}, {Zhao}, {Zhao}, {Zheng}, {Zhou}, \& {Insight-HMXT
  Collaboration}}]{2021ApJ...919...92B}
{Bu}, Q.~C., {Zhang}, S.~N., {Santangelo}, A., {et~al.} 2021, \apj, 919, 92,
  \dodoi{10.3847/1538-4357/ac11f5}

\bibitem[{{Cadolle Bel} {et~al.}(2006){Cadolle Bel}, {Sizun}, {Goldwurm},
  {Rodriguez}, {Laurent}, {Zdziarski}, {Foschini}, {Goldoni}, {Gouiff{\`e}s},
  {Malzac}, {Jourdain}, \& {Roques}}]{2006A&A...446..591C}
{Cadolle Bel}, M., {Sizun}, P., {Goldwurm}, A., {et~al.} 2006, \aap, 446, 591,
  \dodoi{10.1051/0004-6361:20053068}

\bibitem[{{Cangemi} {et~al.}(2023){Cangemi}, {Rodriguez}, {Belloni},
  {Gouiff{\`e}s}, {Grinberg}, {Laurent}, {Petrucci}, \&
  {Wilms}}]{2023A&A...669A..65C}
{Cangemi}, F., {Rodriguez}, J., {Belloni}, T., {et~al.} 2023, \aap, 669, A65,
  \dodoi{10.1051/0004-6361/202243564}

\bibitem[{{Cao} {et~al.}(2020){Cao}, {Jiang}, {Meng}, {Zhang}, {Luo}, {Yang},
  {Zhang}, {Gu}, {Sun}, {Liu}, {Yang}, {Li}, {Tan}, {Liu}, {Du}, {Lu}, {Xu},
  {Guan}, {Zhang}, {Wang}, {Li}, {Zhang}, {Wen}, {Qu}, {Song}, {Li}, {Ge},
  {Zhou}, {Xiong}, {Zhang}, {Zhang}, {Cheng}, {Zhang}, {Li}, {Liang}, {Gao},
  {Yang}, {Liu}, {Liu}, {Yang}, \& {Zhang}}]{2020SCPMA..6349504C}
{Cao}, X., {Jiang}, W., {Meng}, B., {et~al.} 2020, SCPMA, 63, 249504,
  \dodoi{10.1007/s11433-019-1506-1}

\bibitem[{{Casella} {et~al.}(2004){Casella}, {Belloni}, {Homan}, \&
  {Stella}}]{2004A&A...426..587C}
{Casella}, P., {Belloni}, T., {Homan}, J., \& {Stella}, L. 2004, \aap, 426,
  587, \dodoi{10.1051/0004-6361:20041231}

\bibitem[{{Castro-Tirado} {et~al.}(2023){Castro-Tirado}, {Sanchez-Ramirez},
  {Caballero-Garcia}, {Perez-Garcia}, {Fernandez-Garcia}, {Guziy}, {Hu},
  {Blazek}, {Hermelo}, {Pinter}, {Meintjes}, {van Heerden}, {Martin-Carrillo},
  {Hanlon}, {Hiriart}, {Lee}, {Carrasco-Garcia}, {Park}, {Gritsevich},
  {Castellon}, {Perez del Pulgar}, \& {Reina}}]{2023ATel16208....1C}
{Castro-Tirado}, A.~J., {Sanchez-Ramirez}, R., {Caballero-Garcia}, M.~D.,
  {et~al.} 2023, The Astronomer's Telegram, 16208, 1

\bibitem[{{Chen} {et~al.}(2020){Chen}, {Cui}, {Li}, {Wang}, {Xu}, {Lu}, {Wang},
  {Chen}, {Han}, {Hu}, {Zhang}, {Huo}, {Yang}, {Li}, {Lu}, {Zhang}, {Li},
  {Zhang}, {Xiong}, {Zhang}, {Xue}, {Zhao}, {Zhu}, {Zhu}, {Liu}, {Yang}, \&
  {Zhang}}]{2020SCPMA..6349505C}
{Chen}, Y., {Cui}, W., {Li}, W., {et~al.} 2020, SCPMA, 63, 249505,
  \dodoi{10.1007/s11433-019-1469-5}

\bibitem[{{Chen} {et~al.}(2018){Chen}, {Zhang}, {Qu}, {Zhang}, {Ji}, {Kong},
  {Cao}, {Chang}, {Chen}, {Chen}, {Chen}, {Chen}, {Chen}, {Cui}, {Cui}, {Deng},
  {Dong}, {Du}, {Fu}, {Gao}, {Gao}, {Gao}, {Ge}, {Gu}, {Guan}, {Guo}, {Han},
  {Hu}, {Huang}, {Huo}, {Jia}, {Jiang}, {Jiang}, {Jin}, {Jin}, {Li}, {Li},
  {Li}, {Li}, {Li}, {Li}, {Li}, {Li}, {Li}, {Li}, {Li}, {Li}, {Liang}, {Liao},
  {Liu}, {Liu}, {Liu}, {Liu}, {Liu}, {Liu}, {Liu}, {Lu}, {Lu}, {Lu}, {Luo},
  {Ma}, {Meng}, {Nang}, {Nie}, {Ou}, {Sai}, {Song}, {Sun}, {Tan}, {Tao}, {Tao},
  {Tuo}, {Wang}, {Wang}, {Wang}, {Wang}, {Wang}, {Wen}, {Wu}, {Wu}, {Xiao},
  {Xiong}, {Xu}, {Xu}, {Yan}, {Yang}, {Yang}, {Yang}, {Zhang}, {Zhang},
  {Zhang}, {Zhang}, {Zhang}, {Zhang}, {Zhang}, {Zhang}, {Zhang}, {Zhang},
  {Zhang}, {Zhang}, {Zhang}, {Zhang}, {Zhang}, {Zhang}, {Zhang}, {Zhao},
  {Zhao}, {Zhao}, {Zheng}, {Zhu}, {Zhu}, \& {Zou}}]{2018ApJ...864L..30C}
{Chen}, Y.~P., {Zhang}, S., {Qu}, J.~L., {et~al.} 2018, \apjl, 864, L30,
  \dodoi{10.3847/2041-8213/aadc0e}

\bibitem[{{Garc{\'\i}a} {et~al.}(2014){Garc{\'\i}a}, {Dauser}, {Lohfink},
  {Kallman}, {Steiner}, {McClintock}, {Brenneman}, {Wilms}, {Eikmann},
  {Reynolds}, \& {Tombesi}}]{2014ApJ...782...76G}
{Garc{\'\i}a}, J., {Dauser}, T., {Lohfink}, A., {et~al.} 2014, \apj, 782, 76,
  \dodoi{10.1088/0004-637X/782/2/76}

\bibitem[{{Huang} {et~al.}(2018){Huang}, {Qu}, {Zhang}, {Bu}, {Chen}, {Tao},
  {Zhang}, {Lu}, {Li}, {Song}, {Xu}, {Cao}, {Chen}, {Liu}, {Chang}, {Yu},
  {Weng}, {Hou}, {Kong}, {Xie}, {Zhang}, {ZHOU}, {Chang}, {Chen}, {Chen},
  {Chen}, {Chen}, {Cui}, {Cui}, {Deng}, {Dong}, {Du}, {Fu}, {Gao}, {Gao},
  {Gao}, {Ge}, {Gu}, {Guan}, {Gungor}, {Guo}, {Han}, {Hu}, {Huo}, {Ji}, {Jia},
  {Jiang}, {Jiang}, {Jin}, {Jin}, {Li}, {Li}, {Li}, {Li}, {Li}, {Li}, {Li},
  {Li}, {Li}, {Li}, {Li}, {Liang}, {Liao}, {Liu}, {Liu}, {Liu}, {Liu}, {Liu},
  {Liu}, {Lu}, {Lu}, {Luo}, {Ma}, {Meng}, {Nang}, {Nie}, {Ou}, {Sai}, {Shang},
  {Sun}, {Tan}, {Tao}, {Tuo}, {Wang}, {Wang}, {Wang}, {Wang}, {Wang}, {Wen},
  {Wu}, {Wu}, {Xiao}, {Xiong}, {Xu}, {Yan}, {Yang}, {Yang}, {Yang}, {Zhang},
  {Zhang}, {Zhang}, {Zhang}, {Zhang}, {Zhang}, {Zhang}, {Zhang}, {Zhang},
  {Zhang}, {Zhang}, {Zhang}, {Zhang}, {Zhang}, {Zhang}, {Zhang}, {Zhang},
  {Zhang}, {Zhao}, {Zhao}, {Zhao}, {Zheng}, {Zhu}, {Zhu}, {Zou}, \&
  {Insight-HXMT Collaboration}}]{2018ApJ...866..122H}
{Huang}, Y., {Qu}, J.~L., {Zhang}, S.~N., {et~al.} 2018, \apj, 866, 122,
  \dodoi{10.3847/1538-4357/aade4c}

\bibitem[{{Ingram} {et~al.}(2009){Ingram}, {Done}, \&
  {Fragile}}]{2009MNRAS.397L.101I}
{Ingram}, A., {Done}, C., \& {Fragile}, P.~C. 2009, \mnras, 397, L101,
  \dodoi{10.1111/j.1745-3933.2009.00693.x}

\bibitem[{{Ingram} {et~al.}(2023){Ingram}, {Bollemeijer}, {Veledina},
  {Dovciak}, {Poutanen}, {Egron}, {Russell}, {Trushkin}, {Negro}, {Ratheesh},
  {Capitanio}, {Connors}, {Neilsen}, {Kraus}, {Noemi Iacolina}, {Pellizzoni},
  {Pilia}, {Carotenuto}, {Matt}, {Mastroserio}, {Kaaret}, {Bianchi}, {Garcia},
  {Bachetti}, {Wu}, {Costa}, {Ewing}, {Kravtsov}, {Krawczynski}, {Loktev},
  {Marinucci}, {Marra}, {Mikusincova}, {Nathan}, {Parra}, {Petrucci},
  {Righini}, {Soffitta}, {Steiner}, {Svoboda}, {Tombesi}, {Tugliani}, {Ursini},
  {Yang}, {Zane}, {Zhang}, {Agudo}, {Antonelli}, {Baldini}, {Baumgartner},
  {Bellazzini}, {Bongiorno}, {Bonino}, {Brez}, {Bucciantini}, {Castellano},
  {Cavazzuti}, {Chen}, {Ciprini}, {De Rosa}, {Del Monte}, {Di Gesu}, {Di
  Lalla}, {Di Marco}, {Donnarumma}, {Doroshenko}, {Ehlert}, {Enoto},
  {Evangelista}, {Fabiani}, {Ferrazzoli}, {Gunji}, {Hayashida}, {Heyl},
  {Iwakiri}, {Jorstad}, {Karas}, {Kislat}, {Kitaguchi}, {Kolodziejczak}, {La
  Monaca}, {Latronico}, {Liodakis}, {Maldera}, {Manfreda}, {Marin}, {Marscher},
  {Marshall}, {Massaro}, {Mitsuishi}, {Mizuno}, {Muleri}, {Ng}, {O'Dell},
  {Omodei}, {Oppedisano}, {Papitto}, {Pavlov}, {Peirson}, {Perri},
  {Pesce-Rollins}, {Possenti}, {Puccetti}, {Ramsey}, {Rankin}, {Roberts},
  {Romani}, {Sgro}, {Slane}, {Spandre}, {Swartz}, {Tamagawa}, {Tavecchio},
  {Taverna}, {Tawara}, {Tennant}, {Thomas}, {Trois}, {Tsygankov}, {Turolla},
  {Vink}, {Weisskopf}, \& {Xie}}]{2023arXiv231105497I}
{Ingram}, A., {Bollemeijer}, N., {Veledina}, A., {et~al.} 2023, arXiv e-prints,
  arXiv:2311.05497, \dodoi{10.48550/arXiv.2311.05497}

\bibitem[{{Ingram} \& {Motta}(2019)}]{Ingram2019}
{Ingram}, A.~R., \& {Motta}, S.~E. 2019, \nar, 85, 101524,
  \dodoi{10.1016/j.newar.2020.101524}

\bibitem[{{Jourdain} {et~al.}(2014){Jourdain}, {Roques}, \&
  {Chauvin}}]{2014ApJ...789...26J}
{Jourdain}, E., {Roques}, J.~P., \& {Chauvin}, M. 2014, \apj, 789, 26,
  \dodoi{10.1088/0004-637X/789/1/26}

\bibitem[{{Jourdain} {et~al.}(2012){Jourdain}, {Roques}, {Chauvin}, \&
  {Clark}}]{2012ApJ...761...27J}
{Jourdain}, E., {Roques}, J.~P., {Chauvin}, M., \& {Clark}, D.~J. 2012, \apj,
  761, 27, \dodoi{10.1088/0004-637X/761/1/27}

\bibitem[{{Kantzas} {et~al.}(2021){Kantzas}, {Markoff}, {Beuchert}, {Lucchini},
  {Chhotray}, {Ceccobello}, {Tetarenko}, {Miller-Jones}, {Bremer}, {Garcia},
  {Grinberg}, {Uttley}, \& {Wilms}}]{2021MNRAS.500.2112K}
{Kantzas}, D., {Markoff}, S., {Beuchert}, T., {et~al.} 2021, \mnras, 500, 2112,
  \dodoi{10.1093/mnras/staa3349}

\bibitem[{{Kong} {et~al.}(2020){Kong}, {Zhang}, {Chen}, {Ji}, {Zhang}, {Yang},
  {Tao}, {Ma}, {Qu}, {Lu}, {Bu}, {Chen}, {Song}, {Li}, {Xu}, {Cao}, {Chen},
  {Liu}, {Cai}, {Chang}, {Chen}, {Chen}, {Chen}, {Cui}, {Cui}, {Deng}, {Dong},
  {Du}, {Fu}, {Gao}, {Gao}, {Gao}, {Ge}, {Gu}, {Guan}, {Guo}, {Han}, {Huang},
  {Huo}, {Jia}, {Jiang}, {Jiang}, {Jin}, {Li}, {Li}, {Li}, {Li}, {Li}, {Li},
  {Li}, {Li}, {Li}, {Li}, {Liang}, {Liao}, {Liu}, {Liu}, {Liu}, {Liu}, {Liu},
  {Liu}, {Lu}, {Lu}, {Luo}, {Luo}, {Meng}, {Nang}, {Nie}, {Ou}, {Ren}, {Sai},
  {Song}, {Sun}, {Tan}, {Tuo}, {Wang}, {Wang}, {Wang}, {Wang}, {Wang}, {Wang},
  {Wen}, {Wu}, {Wu}, {Wu}, {Xiao}, {Xiao}, {Xiong}, {Xu}, {Yang}, {Yang},
  {Yang}, {Yi}, {You}, {Zhang}, {Zhang}, {Zhang}, {Zhang}, {Zhang}, {Zhang},
  {Zhang}, {Zhang}, {Zhang}, {Zhang}, {Zhang}, {Zhang}, {Zhang}, {Zhang},
  {Zhang}, {Zhang}, {Zhang}, {Zhao}, {Zhao}, {Zheng}, {Zheng}, {Zhou}, {Zhou},
  {Zhu}, {Zhu}, \& {Insight-HXMT Collaboration}}]{2020JHEAp..25...29K}
{Kong}, L.~D., {Zhang}, S., {Chen}, Y.~P., {et~al.} 2020, Journal of High
  Energy Astrophysics, 25, 29, \dodoi{10.1016/j.jheap.2020.01.003}

\bibitem[{{Krawczynski} {et~al.}(2022){Krawczynski}, {Muleri}, {Dov{\v{c}}iak},
  {Veledina}, {Rodriguez Cavero}, {Svoboda}, {Ingram}, {Matt}, {Garcia},
  {Loktev}, {Negro}, {Poutanen}, {Kitaguchi}, {Podgorn{\'y}}, {Rankin},
  {Zhang}, {Berdyugin}, {Berdyugina}, {Bianchi}, {Blinov}, {Capitanio}, {Di
  Lalla}, {Draghis}, {Fabiani}, {Kagitani}, {Kravtsov}, {Kiehlmann},
  {Latronico}, {Lutovinov}, {Mandarakas}, {Marin}, {Marinucci}, {Miller},
  {Mizuno}, {Molkov}, {Omodei}, {Petrucci}, {Ratheesh}, {Sakanoi}, {Semena},
  {Skalidis}, {Soffitta}, {Tennant}, {Thalhammer}, {Tombesi}, {Weisskopf},
  {Wilms}, {Zhang}, {Agudo}, {Antonelli}, {Bachetti}, {Baldini}, {Baumgartner},
  {Bellazzini}, {Bongiorno}, {Bonino}, {Brez}, {Bucciantini}, {Castellano},
  {Cavazzuti}, {Ciprini}, {Costa}, {De Rosa}, {Del Monte}, {Di Gesu}, {Di
  Marco}, {Donnarumma}, {Doroshenko}, {Ehlert}, {Enoto}, {Evangelista},
  {Ferrazzoli}, {Gunji}, {Hayashida}, {Heyl}, {Iwakiri}, {Jorstad}, {Karas},
  {Kolodziejczak}, {La Monaca}, {Liodakis}, {Maldera}, {Manfreda}, {Marscher},
  {Marshall}, {Mitsuishi}, {Ng}, {O{\textquoteright}Dell}, {Oppedisano},
  {Papitto}, {Pavlov}, {Peirson}, {Perri}, {Pesce-Rollins}, {Pilia},
  {Possenti}, {Puccetti}, {Ramsey}, {Romani}, {Sgr{\`o}}, {Slane}, {Spandre},
  {Tamagawa}, {Tavecchio}, {Taverna}, {Tawara}, {Thomas}, {Trois}, {Tsygankov},
  {Turolla}, {Vink}, {Wu}, {Xie}, \& {Zane}}]{2022Sci...378..650K}
{Krawczynski}, H., {Muleri}, F., {Dov{\v{c}}iak}, M., {et~al.} 2022, Science,
  378, 650, \dodoi{10.1126/science.add5399}

\bibitem[{{Laurent} {et~al.}(2011){Laurent}, {Rodriguez}, {Wilms}, {Cadolle
  Bel}, {Pottschmidt}, \& {Grinberg}}]{2011Sci...332..438L}
{Laurent}, P., {Rodriguez}, J., {Wilms}, J., {et~al.} 2011, Science, 332, 438,
  \dodoi{10.1126/science.1200848}

\bibitem[{{Li} {et~al.}(2013){Li}, {Zhang}, {Qu}, {Gao}, {Zhao}, {Huang}, \&
  {Song}}]{2013MNRAS.433..412L}
{Li}, Z.~B., {Zhang}, S., {Qu}, J.~L., {et~al.} 2013, \mnras, 433, 412,
  \dodoi{10.1093/mnras/stt737}

\bibitem[{{Liu} {et~al.}(2020){Liu}, {Zhang}, {Li}, {Lu}, {Chang}, {Li},
  {Zhang}, {Jin}, {Yu}, {Zhang}, {Fu}, {Chen}, {Ji}, {Xu}, {Deng}, {Shang},
  {Liu}, {Lu}, {Zhang}, {Dong}, {Li}, {Wu}, {Li}, {Wang}, {Wu}, {Zhang},
  {Zhang}, {Xiong}, {Liu}, {Zhang}, {Liu}, {Yang}, \&
  {Zhang}}]{2020SCPMA..6349503L}
{Liu}, C., {Zhang}, Y., {Li}, X., {et~al.} 2020, SCPMA, 63, 249503,
  \dodoi{10.1007/s11433-019-1486-x}

\bibitem[{{Liu} {et~al.}(2024){Liu}, {Xu}, {Zhang}, {Yu}, {Huang}, {Tao},
  {Zhang}, {Yang}, {Zhao}, {Qu}, \& {Song}}]{2024arXiv240603834L}
{Liu}, H.-X., {Xu}, Y.-J., {Zhang}, S.-N., {et~al.} 2024, arXiv e-prints,
  arXiv:2406.03834, \dodoi{10.48550/arXiv.2406.03834}

\bibitem[{{Liu} {et~al.}(2023){Liu}, {Li}, {Pan}, {Liu}, {Ling}, {Zhang},
  {Cheng}, {Cui}, {Fan}, {Hu}, {Hu}, {Huang}, {Jin}, {Li}, {Li}, {Li}, {Liu},
  {Sun}, {Wang}, {Wang}, {Wang}, {Wu}, {Xu}, {Xu}, {Yang}, {Zhang}, {Zhang},
  {Zhang}, {Zhang}, {Zhao}, \& {Yuan}}]{2023ATel16210....1L}
{Liu}, H.~Y., {Li}, D.~Y., {Pan}, H.~W., {et~al.} 2023, The Astronomer's
  Telegram, 16210, 1

\bibitem[{{Ma} {et~al.}(2021){Ma}, {Tao}, {Zhang}, {Zhang}, {Bu}, {Ge}, {Chen},
  {Qu}, {Zhang}, {Lu}, {Song}, {Yang}, {Yuan}, {Cai}, {Cao}, {Chang}, {Chen},
  {Chen}, {Chen}, {Chen}, {Chen}, {Cui}, {Cui}, {Deng}, {Dong}, {Du}, {Fu},
  {Gao}, {Gao}, {Gao}, {Gu}, {Guan}, {Guo}, {Han}, {Huang}, {Huo}, {Ji}, {Jia},
  {Jiang}, {Jiang}, {Jin}, {Jin}, {Kong}, {Li}, {Li}, {Li}, {Li}, {Li}, {Li},
  {Li}, {Li}, {Li}, {Li}, {Li}, {Liang}, {Liao}, {Liu}, {Liu}, {Liu}, {Liu},
  {Liu}, {Liu}, {Lu}, {Lu}, {Luo}, {Luo}, {Meng}, {Nang}, {Nie}, {Ou}, {Sai},
  {Shang}, {Song}, {Sun}, {Tan}, {Tuo}, {Wang}, {Wang}, {Wang}, {Wang}, {Wang},
  {Wang}, {Wen}, {Wu}, {Wu}, {Wu}, {Xiao}, {Xiao}, {Xie}, {Xiong}, {Xu}, {Xu},
  {Yang}, {Yang}, {Yang}, {Yi}, {Yin}, {You}, {Zhang}, {Zhang}, {Zhang},
  {Zhang}, {Zhang}, {Zhang}, {Zhang}, {Zhang}, {Zhang}, {Zhang}, {Zhang},
  {Zhang}, {Zhang}, {Zhang}, {Zhang}, {Zhang}, {Zhao}, {Zhao}, {Zheng}, {Zhou},
  {Zhou}, {Zhu}, {Zhu}, \& {Zhuang}}]{2021NatAs...5...94M}
{Ma}, X., {Tao}, L., {Zhang}, S.-N., {et~al.} 2021, Nature Astronomy, 5, 94,
  \dodoi{10.1038/s41550-020-1192-2}

\bibitem[{{Ma} {et~al.}(2023){Ma}, {Zhang}, {Tao}, {Bu}, {Qu}, {Zhang}, {Zhou},
  {Huang}, {Jia}, {Song}, {Zhang}, {Ge}, {Liu}, {Yang}, {Yu}, \&
  {Yorgancioglu}}]{2023ApJ...948..116M}
{Ma}, X., {Zhang}, L., {Tao}, L., {et~al.} 2023, \apj, 948, 116,
  \dodoi{10.3847/1538-4357/acc4c3}

\bibitem[{{Markoff} {et~al.}(2005){Markoff}, {Nowak}, \&
  {Wilms}}]{2005ApJ...635.1203M}
{Markoff}, S., {Nowak}, M.~A., \& {Wilms}, J. 2005, \apj, 635, 1203,
  \dodoi{10.1086/497628}

\bibitem[{{M{\'e}ndez} {et~al.}(2022){M{\'e}ndez}, {Karpouzas}, {Garc{\'\i}a},
  {Zhang}, {Zhang}, {Belloni}, \& {Altamirano}}]{2022NatAs...6..577M}
{M{\'e}ndez}, M., {Karpouzas}, K., {Garc{\'\i}a}, F., {et~al.} 2022, Nature
  Astronomy, 6, 577, \dodoi{10.1038/s41550-022-01617-y}

\bibitem[{{M{\'e}ndez} {et~al.}(2024){M{\'e}ndez}, {Peirano}, {Garc{\'\i}a},
  {Belloni}, {Altamirano}, \& {Alabarta}}]{2024MNRAS.527.9405M}
{M{\'e}ndez}, M., {Peirano}, V., {Garc{\'\i}a}, F., {et~al.} 2024, \mnras, 527,
  9405, \dodoi{10.1093/mnras/stad3786}

\bibitem[{{Miller-Jones} {et~al.}(2023){Miller-Jones}, {Sivakoff}, {Bahramian},
  \& {Russell}}]{2023ATel16211....1M}
{Miller-Jones}, J.~C.~A., {Sivakoff}, G.~R., {Bahramian}, A., \& {Russell},
  T.~D. 2023, The Astronomer's Telegram, 16211, 1

\bibitem[{{Mitsuda} {et~al.}(1984){Mitsuda}, {Inoue}, {Koyama}, {Makishima},
  {Matsuoka}, {Ogawara}, {Shibazaki}, {Suzuki}, {Tanaka}, \&
  {Hirano}}]{1984PASJ...36..741M}
{Mitsuda}, K., {Inoue}, H., {Koyama}, K., {et~al.} 1984, \pasj, 36, 741

\bibitem[{{Miyamoto} {et~al.}(1991){Miyamoto}, {Kimura}, {Kitamoto}, {Dotani},
  \& {Ebisawa}}]{1991ApJ...383..784M}
{Miyamoto}, S., {Kimura}, K., {Kitamoto}, S., {Dotani}, T., \& {Ebisawa}, K.
  1991, \apj, 383, 784, \dodoi{10.1086/170837}

\bibitem[{{Nakajima} {et~al.}(2023){Nakajima}, {Negoro}, {Serino}, {Mihara},
  {Kobayashi}, {Tanaka}, {Soejima}, {Kudo}, {Kawamuro}, {Yamada}, {Tamagawa},
  {Kawai}, {Matsuoka}, {Sakamoto}, {Sugita}, {Hiramatsu}, {Nishikawa},
  {Yoshida}, {Tsuboi}, {Urabe}, {Nawa}, {Nemoto}, {Shidatsu}, {Takahashi},
  {Niwano}, {Sato}, {Higuchi}, {Yatsu}, {Nakahira}, {Ueno}, {Tomida},
  {Ishikawa}, {Ogawa}, {Kurihara}, {Ueda}, {Setoguchi}, {Yoshitake},
  {Nakatani}, {Yamauchi}, {Hagiwara}, {Umeki}, {Otsuki}, {Yamaoka}, {Kawakubo},
  {Sugizaki}, \& {Iwakiri}}]{2023ATel16206....1N}
{Nakajima}, M., {Negoro}, H., {Serino}, M., {et~al.} 2023, The Astronomer's
  Telegram, 16206, 1

\bibitem[{{Nandi} {et~al.}(2024){Nandi}, {Das}, {Majumder}, {Katoch}, {Antia},
  \& {Shah}}]{2024arXiv240417160N}
{Nandi}, A., {Das}, S., {Majumder}, S., {et~al.} 2024, arXiv e-prints,
  arXiv:2404.17160, \dodoi{10.48550/arXiv.2404.17160}

\bibitem[{{Negoro} {et~al.}(2023){Negoro}, {Serino}, {Nakajima}, {Kobayashi},
  {Tanaka}, {Soejima}, {Kudo}, {Mihara}, {Kawamuro}, {Yamada}, {Tamagawa},
  {Kawai}, {Matsuoka}, {Sakamoto}, {Sugita}, {Hiramatsu}, {Nishikawa},
  {Yoshida}, {Tsuboi}, {Urabe}, {Nawa}, {Nemoto}, {Shidatsu}, {Takahashi},
  {Niwano}, {Sato}, {Higuchi}, {Yatsu}, {Nakahira}, {Ueno}, {Tomida},
  {Ishikawa}, {Ogawa}, {Kurihara}, {Ueda}, {Setoguchi}, {Yoshitake},
  {Nakatani}, {Yamauchi}, {Hagiwara}, {Umeki}, {Otsuki}, {Yamaoka}, {Kawakubo},
  {Sugizaki}, \& {Iwakiri}}]{2023ATel16205....1N}
{Negoro}, H., {Serino}, M., {Nakajima}, M., {et~al.} 2023, The Astronomer's
  Telegram, 16205, 1

\bibitem[{{Peng} {et~al.}(2024){Peng}, {Zhang}, {Shui}, {Zhang}, {Kong},
  {Chen}, {Wang}, {Ji}, {Qu}, {Tao}, {Ge}, {Chang}, {Li}, {Li}, {Yu}, \&
  {Yan}}]{2024ApJ...960L..17P}
{Peng}, J.-Q., {Zhang}, S., {Shui}, Q.-C., {et~al.} 2024, \apjl, 960, L17,
  \dodoi{10.3847/2041-8213/ad17ca}

\bibitem[{{Podgorn{\'y}} {et~al.}(2024){Podgorn{\'y}}, {Svoboda},
  {Dov{\v{c}}iak}, {Veledina}, {Poutanen}, {Kaaret}, {Bianchi}, {Ingram},
  {Capitanio}, {Datta}, {Egron}, {Krawczynski}, {Matt}, {Muleri}, {Petrucci},
  {Russell}, {Steiner}, {Bollemeijer}, {Brigitte}, {Emami}, {Garc{\'\i}a},
  {Hu}, {Iacolina}, {Kravtsov}, {Marra}, {Mastroserio}, {Mu{\~n}oz-Darias},
  {Nathan}, {Negro}, {Ratheesh}, {Rodriguez Cavero}, {Taverna}, {Tombesi},
  {Yang}, {Zhang}, \& {Zhang}}]{2024arXiv240419601P}
{Podgorn{\'y}}, J., {Svoboda}, J., {Dov{\v{c}}iak}, M., {et~al.} 2024, arXiv
  e-prints, arXiv:2404.19601, \dodoi{10.48550/arXiv.2404.19601}

\bibitem[{{Rodriguez} {et~al.}(2004){Rodriguez}, {Corbel}, {Hannikainen},
  {Belloni}, {Paizis}, \& {Vilhu}}]{2004ApJ...615..416R}
{Rodriguez}, J., {Corbel}, S., {Hannikainen}, D.~C., {et~al.} 2004, \apj, 615,
  416, \dodoi{10.1086/423978}

\bibitem[{{Rodriguez} {et~al.}(2015){Rodriguez}, {Grinberg}, {Laurent},
  {Cadolle Bel}, {Pottschmidt}, {Pooley}, {Bodaghee}, {Wilms}, \&
  {Gouiff{\`e}s}}]{2015ApJ...807...17R}
{Rodriguez}, J., {Grinberg}, V., {Laurent}, P., {et~al.} 2015, \apj, 807, 17,
  \dodoi{10.1088/0004-637X/807/1/17}

\bibitem[{{Sunyaev} {et~al.}(2023){Sunyaev}, {Mereminskiy}, {Molkov}, {Semena},
  {Arefiev}, {Krivonos}, {Levin}, {Lutovinov}, {Shtykovsky}, \&
  {Tkachenko}}]{2023ATel16217....1S}
{Sunyaev}, R.~A., {Mereminskiy}, I.~A., {Molkov}, S.~V., {et~al.} 2023, The
  Astronomer's Telegram, 16217, 1

\bibitem[{{Svoboda} {et~al.}(2024){Svoboda}, {Dov{\v{c}}iak}, {Steiner},
  {Kaaret}, {Podgorn{\'y}}, {Poutanen}, {Veledina}, {Muleri}, {Taverna},
  {Krawczynski}, {Brigitte}, {Ranjan Datta}, {Bianchi}, {Castro Segura},
  {Garc{\'\i}a}, {Ingram}, {Matt}, {Mu{\~n}oz-Darias}, {Nathan}, {Weisskopf},
  {Altamirano}, {Baldini}, {Bollemeijer}, {Capitanio}, {Egron}, {Emami}, {Hu},
  {Marra}, {Mastroserio}, {Negro}, {Petrucci}, {Ratheesh}, {Rodriguez Cavero},
  {Soffitta}, {Tombesi}, {Yang}, \& {Zhang}}]{2024arXiv240304689S}
{Svoboda}, J., {Dov{\v{c}}iak}, M., {Steiner}, J.~F., {et~al.} 2024, arXiv
  e-prints, arXiv:2403.04689, \dodoi{10.48550/arXiv.2403.04689}

\bibitem[{{Veledina} {et~al.}(2023){Veledina}, {Muleri}, {Dov{\v{c}}iak},
  {Poutanen}, {Ratheesh}, {Capitanio}, {Matt}, {Soffitta}, {Tennant}, {Negro},
  {Kaaret}, {Costa}, {Ingram}, {Svoboda}, {Krawczynski}, {Bianchi}, {Steiner},
  {Garc{\'\i}a}, {Kravtsov}, {Nitindala}, {Ewing}, {Mastroserio}, {Marinucci},
  {Ursini}, {Tombesi}, {Tsygankov}, {Yang}, {Weisskopf}, {Trushkin}, {Egron},
  {Iacolina}, {Pilia}, {Marra}, {Miku{\v{s}}incov{\'a}}, {Nathan}, {Parra},
  {Petrucci}, {Podgorn{\'y}}, {Tugliani}, {Zane}, {Zhang}, {Agudo},
  {Antonelli}, {Bachetti}, {Baldini}, {Baumgartner}, {Bellazzini}, {Bongiorno},
  {Bonino}, {Brez}, {Bucciantini}, {Castellano}, {Cavazzuti}, {Chen},
  {Ciprini}, {De Rosa}, {Del Monte}, {Di Gesu}, {Di Lalla}, {Di Marco},
  {Donnarumma}, {Doroshenko}, {Ehlert}, {Enoto}, {Evangelista}, {Fabiani},
  {Ferrazzoli}, {Gunji}, {Hayashida}, {Heyl}, {Iwakiri}, {Jorstad}, {Karas},
  {Kislat}, {Kitaguchi}, {Kolodziejczak}, {La Monaca}, {Latronico}, {Liodakis},
  {Maldera}, {Manfreda}, {Marin}, {Marscher}, {Marshall}, {Massaro},
  {Mitsuishi}, {Mizuno}, {Ng}, {O'Dell}, {Omodei}, {Oppedisano}, {Papitto},
  {Pavlov}, {Peirson}, {Perri}, {Pesce-Rollins}, {Possenti}, {Puccetti},
  {Ramsey}, {Rankin}, {Roberts}, {Romani}, {Sgr{\`o}}, {Slane}, {Spandre},
  {Swartz}, {Tamagawa}, {Tavecchio}, {Taverna}, {Tawara}, {Thomas}, {Trois},
  {Turolla}, {Vink}, {Wu}, \& {Xie}}]{2023ApJ...958L..16V}
{Veledina}, A., {Muleri}, F., {Dov{\v{c}}iak}, M., {et~al.} 2023, \apjl, 958,
  L16, \dodoi{10.3847/2041-8213/ad0781}

\bibitem[{{Wang} \& {Bellm}(2023)}]{2023ATel16209....1W}
{Wang}, Y.~D., \& {Bellm}, E.~C. 2023, The Astronomer's Telegram, 16209, 1

\bibitem[{{Wilms} {et~al.}(2000){Wilms}, {Allen}, \&
  {McCray}}]{2000ApJ...542..914W}
{Wilms}, J., {Allen}, A., \& {McCray}, R. 2000, \apj, 542, 914,
  \dodoi{10.1086/317016}

\bibitem[{{Yadav} {et~al.}(2016){Yadav}, {Misra}, {Verdhan Chauhan}, {Agrawal},
  {Antia}, {Pahari}, {Dedhia}, {Katoch}, {Madhwani}, {Manchanda}, {Paul},
  {Shah}, \& {Ishwara-Chandra}}]{2016ApJ...833...27Y}
{Yadav}, J.~S., {Misra}, R., {Verdhan Chauhan}, J., {et~al.} 2016, \apj, 833,
  27, \dodoi{10.3847/0004-637X/833/1/27}

\bibitem[{{Yan} {et~al.}(2013){Yan}, {Ding}, {Wang}, {Qu}, \&
  {Song}}]{2013MNRAS.434...59Y}
{Yan}, S.-P., {Ding}, G.-Q., {Wang}, N., {Qu}, J.-L., \& {Song}, L.-M. 2013,
  \mnras, 434, 59, \dodoi{10.1093/mnras/stt968}

\bibitem[{{Yan} {et~al.}(2012){Yan}, {Qu}, {Ding}, {Han}, {Song}, {Zhang},
  {Yin}, {Zhang}, \& {Wang}}]{2012Ap&SS.337..137Y}
{Yan}, S.-P., {Qu}, J.-L., {Ding}, G.-Q., {et~al.} 2012, \apss, 337, 137,
  \dodoi{10.1007/s10509-011-0804-9}

\bibitem[{{Yang} {et~al.}(2022){Yang}, {Zhang}, {Huang}, {Bu}, {Zhang}, {Liu},
  {Yu}, {Wang}, {Zhao}, {Tao}, {Qu}, {Zhang}, {Zhang}, {Song}, {Lu}, {Cao},
  {Chen}, {Cai}, {Chang}, {Chen}, {Chen}, {Chen}, {Chen}, {Cui}, {Ding}, {Du},
  {Gao}, {Gao}, {Ge}, {Gu}, {Guan}, {Guo}, {Han}, {Huo}, {Jia}, {Jiang}, {Jin},
  {Kong}, {Li}, {Li}, {Li}, {Li}, {Li}, {Li}, {Li}, {Lin}, {Liu}, {Li}, {Li},
  {Liang}, {Liao}, {Liu}, {Liu}, {Lu}, {Luo}, {Luo}, {Ma}, {Ma}, {Ma}, {Meng},
  {Nang}, {Nie}, {Ou}, {Ren}, {Sai}, {Song}, {Sun}, {Tan}, {Tuo}, {Wang},
  {Wang}, {Wang}, {Wang}, {Wang}, {Wen}, {Wu}, {Wu}, {Wu}, {Xiao}, {Xu},
  {Xiong}, {Yang}, {Yang}, {Yi}, {Yin}, {You}, {Zhang}, {Zhang}, {Zhang},
  {Zhang}, {Zhang}, {Zhang}, {Zhang}, {Zhang}, {Zhang}, {Zhang}, {Zhao},
  {Zhao}, {Zheng}, \& {Zhou}}]{2022ApJ...937...33Y}
{Yang}, Z.-x., {Zhang}, L., {Huang}, Y., {et~al.} 2022, \apj, 937, 33,
  \dodoi{10.3847/1538-4357/ac84d6}

\bibitem[{{Yu} {et~al.}(2024){Yu}, {Bu}, {Zhang}, {Liu}, {Zhang}, {Ducci},
  {Tao}, {Santangelo}, {Doroshenko}, {Huang}, {Yang}, \&
  {Qu}}]{2024MNRAS.tmp..851Y}
{Yu}, W., {Bu}, Q.-C., {Zhang}, S.-N., {et~al.} 2024, \mnras,
  \dodoi{10.1093/mnras/stae835}

\bibitem[{{Zdziarski} {et~al.}(2021){Zdziarski}, {Dzie{\l}ak}, {De Marco},
  {Szanecki}, \& {Nied{\'z}wiecki}}]{2021ApJ...909L...9Z}
{Zdziarski}, A.~A., {Dzie{\l}ak}, M.~A., {De Marco}, B., {Szanecki}, M., \&
  {Nied{\'z}wiecki}, A. 2021, \apjl, 909, L9, \dodoi{10.3847/2041-8213/abe7ef}

\bibitem[{{Zdziarski} {et~al.}(2020){Zdziarski}, {Szanecki}, {Poutanen},
  {Gierli{\'n}ski}, \& {Biernacki}}]{2020MNRAS.492.5234Z}
{Zdziarski}, A.~A., {Szanecki}, M., {Poutanen}, J., {Gierli{\'n}ski}, M., \&
  {Biernacki}, P. 2020, \mnras, 492, 5234, \dodoi{10.1093/mnras/staa159}

\bibitem[{{Zdziarski} {et~al.}(2022){Zdziarski}, {You}, {Szanecki}, {Li}, \&
  {Ge}}]{2022ApJ...928...11Z}
{Zdziarski}, A.~A., {You}, B., {Szanecki}, M., {Li}, X.-B., \& {Ge}, M. 2022,
  \apj, 928, 11, \dodoi{10.3847/1538-4357/ac54a7}

\bibitem[{{Zhang} {et~al.}(2020){Zhang}, {Li}, {Lu}, {Song}, {Xu}, {Liu},
  {Chen}, {Cao}, {Bu}, {Chang}, {Chen}, {Chen}, {Chen}, {Chen}, {Chen}, {Cui},
  {Cui}, {Deng}, {Dong}, {Du}, {Fu}, {Gao}, {Gao}, {Gao}, {Ge}, {Gu}, {Guan},
  {Gungor}, {Guo}, {Han}, {Hu}, {Huang}, {Huo}, {Jia}, {Jiang}, {Jiang}, {Jin},
  {Jin}, {Li}, {Li}, {Li}, {Li}, {Li}, {Li}, {Li}, {Li}, {Li}, {Li}, {Li},
  {Liang}, {Liao}, {Liu}, {Liu}, {Liu}, {Liu}, {Liu}, {Liu}, {Lu}, {Lu}, {Luo},
  {Ma}, {Meng}, {Nang}, {Nie}, {Ou}, {Qu}, {Sai}, {Shang}, {Shen}, {Sun},
  {Tan}, {Tao}, {Tuo}, {Wang}, {Wang}, {Wang}, {Wang}, {Wang}, {Wang}, {Wang},
  {Wen}, {Wu}, {Wu}, {Wu}, {Xiao}, {Xiong}, {Yan}, {Yang}, {Yang}, {Yang},
  {Yi}, {Yuan}, {Zhang}, {Zhang}, {Zhang}, {Zhang}, {Zhang}, {Zhang}, {Zhang},
  {Zhang}, {Zhang}, {Zhang}, {Zhang}, {Zhang}, {Zhang}, {Zhang}, {Zhang},
  {Zhang}, {Zhang}, {Zhang}, {Zhang}, {Zhang}, {Zhao}, {Zhao}, {Zheng}, {Zhou},
  {Zhu}, {Zhu}, {Zhuang}, \& {Insight-HXMT Team}}]{2020SCPMA..6349502Z}
{Zhang}, S.-N., {Li}, T., {Lu}, F., {et~al.} 2020, SCPMA, 63, 249502,
  \dodoi{10.1007/s11433-019-1432-6}

\bibitem[{{Zhang} {et~al.}(2022){Zhang}, {M{\'e}ndez}, {Garc{\'\i}a}, {Zhang},
  {Karpouzas}, {Altamirano}, {Belloni}, {Qu}, {Zhang}, {Tao}, {Zhang}, {Huang},
  {Kong}, {Ma}, {Yu}, {Rawat}, \& {Bellavita}}]{2022MNRAS.512.2686Z}
{Zhang}, Y., {M{\'e}ndez}, M., {Garc{\'\i}a}, F., {et~al.} 2022, \mnras, 512,
  2686, \dodoi{10.1093/mnras/stac690}

\bibitem[{{Zhao} {et~al.}(2024){Zhao}, {Tao}, {Li}, {Zhang}, {Feng}, {Ge},
  {Ji}, {Wang}, {Huang}, {Ma}, {Zhang}, {Qu}, {Xu}, {Zhang}, {Yin}, {Shui},
  {Ma}, {Zhao}, {Li}, {Yang}, {Liu}, \& {Yu}}]{2024ApJ...961L..42Z}
{Zhao}, Q.-C., {Tao}, L., {Li}, H.-C., {et~al.} 2024, \apjl, 961, L42,
  \dodoi{10.3847/2041-8213/ad1e6c}

\bibitem[{{Zhu} \& {Wang}(2024)}]{2024arXiv240509772Z}
{Zhu}, H., \& {Wang}, W. 2024, arXiv e-prints, arXiv:2405.09772,
  \dodoi{10.48550/arXiv.2405.09772}

\end{thebibliography}
\bibliographystyle{aasjournal}



\end{document}